\documentstyle[preprint,prl,aps,epsfig,here]{revtex}
\def\dm2{$\Delta m^2$}
\def\nue{$\nu_e~$}
\def\numu{$\nu_{\mu}~$}
\def\nutau{$\nu_{\tau}~$}
\def\nuone{$\nu_1~$}
\def\nutwo{$\nu_2~$}
\def\nuthree{$\nu_3~$}
\def\mD{$m_D~$}
\def\MR{$M_R~$}

\begin{document}
\draft
\begin{titlepage}
\preprint{\vbox{\baselineskip 10pt{
\hbox{IASSNS-HEP-97/75}
\hbox{SISSA 93/97/EP}
\hbox{November 1997}}}}
\vskip -0.4cm
\title{ \bf Solar Neutrinos and Grand Unification   }
\author{K.S. Babu$^{1)},~$ Q.Y. Liu$^{2)},~$  A.Yu. Smirnov$^{1,3)}$}
%\vglue -0.4cm
\address{1) Institute for Advanced Study,
School of Natural Sciences, Princeton, NJ 08540, USA}
\address{2) Scuola Internazionale Superiore di Studi Avanzati, I-34013
Trieste, Italy}
\address{3) International Centre for Theoretical Physics, I-34100
Trieste, Italy}
\maketitle
\begin{abstract}
\begin{minipage}{5in}
\baselineskip 16pt

We consider the Grand Unification ($GU$) scenario for neutrino masses
which is based on the see-saw mechanism
with the mass of the heaviest right handed (RH) neutrino at the GU-scale:
$M_3 \sim \Lambda_{GU}$, and on the quark-lepton symmetry for  fermions
from the third generation.
The scenario predicts for the light  neutrinos:
$m_3 \sim (2 - 4) \cdot 10^{-3}$ eV  and
$m_2 \sim (0.3 - 3) \cdot 10^{-5}$ eV (in the case of a {\it linear mass
hierarchy} of the RH neutrinos or/and in presence of the Planck scale
suppressed non-renormalizable operators).
It also predicts large $\nu_e - \nu_{\mu}$ mixing:
$~\sin^2 2\theta_{e\mu} \stackrel{_>}{_\sim} 0.2$.
In this scenario  the solar neutrinos ($\nu_{\odot}$)
undergo both the
\nue $\rightarrow$  \nutau  resonance conversion
in the Sun and substantial \nue $\rightarrow$ \numu vacuum oscillations
on the way from the Sun to the Earth.
The interplay of both effects enlarges the  range of
neutrino parameters which solve the $\nu_{\odot}$-problem.
In particular, $\nu_e - \nu_{\tau}$ mixing angle
can be as small as the
corresponding quark mixing:
$\sin^2 2\theta_{e\tau} \geq
(2~-~5) \cdot 10^{-4}$.
The scenario predicts
peculiar (oscillatory) distortion of the boron
neutrino energy spectrum and seasonal variations of
signals. Manifestations of these effects in the Super-Kamiokande
and SNO experiments are studied.

\end{minipage}
\end{abstract}
\thispagestyle{empty}
\end{titlepage}
\newpage

\hsize 16.5truecm
\vsize 24.0truecm
\font\eightrm=cmr8
\setcounter{footnote}{0}
\vglue 1.5cm
\leftline{\bf 1. Introduction}

Small neutrino masses,  in particular, the ones implied
by the solar neutrino data, may be  considered as a possible indication
of Grand Unification ($GU$).  This statement holds in the following
context (which we will call the Grand Unification scenario for
neutrino masses).

1. Neutrino masses are generated by the see-saw mechanism
\cite{SW1,SW2}:
\begin{equation}
{\displaystyle m = -{ m_D^2 \over {M}} = - \frac{h_{D}^{2}v_u^2}{M}
}~,
\label{mSeeSaw}
\end {equation}
where \mD is the Dirac neutrino mass,
$m_D=h_Dv_u$,
$h_D$ is the
neutrino Yukawa coupling, $v_u$ is the VEV of the Higgs doublet $H_u$,
and $M$ is the Majorana mass of the right handed (RH) neutrino.
A direct mass term for the left handed  components may be generated
by the induced VEV of the Higgs triplet of $SU(2)_L$ \cite{SW2}.
We assume that this term,  if it exists, is much smaller than that in
(\ref{mSeeSaw}), at least for the heaviest neutrino.

2. Quark-lepton symmetry is realized for the
third generation of fermions at the
Grand Unification scale, $\Lambda_{GU} = 2\cdot 10^{16}$ GeV,
which is  defined  as the scale of the gauge coupling
unification in the Minimal Supersymmetric Standard Model.
The quark-lepton symmetry
relates the (Dirac type) Yukawa coupling of the neutrino  with
the Yukawa coupling of the top quark at $\Lambda_{GU}$:
\begin{equation}
h_{3D}(\Lambda_{GU})~ = ~h_{t}(\Lambda_{GU}).
\label{mDup}
\end{equation}
At the electroweak scale $h_t(M_{EW})=m_t/{v_u}$, where
$m_t$ is the top quark mass.
The existence of quark-lepton symmetry and the equality
(\ref{mDup}) for the third generation are supported by
the asymptotic mass relation $m_b(\Lambda_{GU}) =
m_\tau(\Lambda_{GU})$  which after renormalization effects results in a
successful prediction of the $b$--quark mass at low energies in
supersymmetric GUT's \cite{btau}.

The boundary condition (\ref{mDup}) is satisfied,
in particular, in GUT's based on  $SO_{10}$.
Furthermore, in
$SO_{10}$--type unification all four Yukawa couplings
in the third generation can be equal:
$h_b = h_\tau = h_t = h_{3D}$.

Exact quark-lepton symmetry is, however, broken for the light generations.
This follows from
the failure of the asymptotic mass relations
$m_s(\Lambda_{GU}) = m_\mu(\Lambda_{GU})$ and $m_d(\Lambda_{GU}) =
m_e(\Lambda_{GU})$  which
are inconsistent with low energy determinations.
One possible origin of this breakdown is non-renormalizable operators
which
are suppressed by $\Lambda_{GU}/M_P$, where $M_P$ is the Planck scale
\cite{nonr}.  These  operators
can contribute significantly to the lighter generation masses while leaving
the third generation relation essentially uncorrected.
Therefore for the lighter generations one can  expect
relations of the type (\ref{mDup}) to hold only as an order of magnitude.
One would also expect the leptonic mixing angles involving
the lighter generations to be different from their quark analogs.

3. The Majorana mass of the RH neutrino from the third generation
is at the $GU$ scale: $M_3 \sim \Lambda_{GU}$. Using this mass,
the boundary condition (\ref{mDup}) and the known value of the  
top quark mass one finds from the see-saw formula
the mass of the third neutrino ($\approx \nu_{\tau}$)
$m_3 \sim$ several $\times 10^{-3}$ eV.
The mass $m_3$ is in the range suggested by the
Mikheyev-Smirnov-Wolfenstein (MSW) solution of the
solar ($\nu_{\odot}$-) neutrino problem \cite{ms}:
\begin{equation}
m ~=~(2 ~- ~4)\cdot 10^{-3}~\rm{eV}
\label{mMSW}
\end {equation}
(barring any degeneracy in the spectrum).
Thus in the $GU$-scenario the
$\nu_{\odot}$-problem  can be solved by
the \nue $\rightarrow$ \nutau resonance
conversion in the Sun
provided the mixing between the first
and the third generations is sufficiently large.
{\it Vice versa},
using the value of the light neutrino mass
(\ref{mMSW}) and the top quark mass
at the electroweak scale along with the boundary condition
(\ref{mDup}) one finds from the
see-saw formula
$M_{3} \sim 10^{16}$ GeV \cite{Lan}.
The approximate equality of the mass scales $\Lambda_{GU}$  and $M_3$
may be considered as a hint for Grand Unification in addition to
the gauge coupling unification and the $b-\tau$ unification.

An alternative solution to the solar neutrino problem is via
\nue $\rightarrow$ \numu resonance conversion.
In this case the boundary condition for the second
generation,
$m_{2D}(\Lambda_{GU}) \approx m_c(\Lambda_{GU})$
($m_c$ is the charm quark mass),
leads to  $M_{2} \sim (10^{10}~-~10^{11})$ GeV
in the intermediate mass scale
(see, {\it e.g.}, \cite{smirnovint}).

Note that in the 2$\nu$-case, it is impossible to distinguish
\nue $\rightarrow$ \numu and \nue $\rightarrow$ \nutau
transitions using the $\nu_\odot$-data only.
Both \numu and \nutau are detected by the neutral current
interactions which are the same (up to higher order corrections) for both
neutrinos. The situation can be different if the mixing of
all three neutrinos is taken into account.
We will show that solar neutrino data alone may
disentangle the two possibilities ($GU$-scale scenario and the
intermediate scale scenario).

A possible  discovery of $\nu$-oscillations by
CHORUS/NOMAD \cite{CHORUSNOMAD} or new experiments like E803 (COSMOS)
\cite{COSMOS} would imply $m_3\sim $
O(1 eV), favoring  the intermediate scale physics.
Also  confirmations of the LSND  result \cite{lsnd}
and oscillation interpretation of the atmospheric
neutrino anomaly
will exclude the simplest version of the
$GU$-scenario.

In this paper we reconsider the Grand Unification
scenario and study its  signatures
in the solar neutrinos. We calculate $M_3$  implied by the
\nue $\rightarrow$ \nutau
solution of the $\nu_\odot$-problem using the
known mass of the top quark and  discuss
possible relations of $M_3$ to $\Lambda_{GU}$ (see sect. 2).
The  \nue - \nutau mixing is considered in sect. 3.
We show that it is quite plausible that
\nue $\rightarrow$ \nutau conversion will be
accompanied by sizable \nue $\rightarrow$ \numu vacuum oscillations
of solar neutrinos (sect. 4).
The interplay of both transitions (sect. 5)
has a number of consequences: it
enlarges the region of neutrino parameters in which one can get
a correct description of the $\nu_{\odot}$-data (sect. 6),
it leads to peculiar distortion of the boron neutrino
spectrum (sect. 7) and to seasonal variations of signals (sect. 8).
Possible manifestations of
these  effects which can be considered as signatures of
the $GU$-scenario
in the  Super-Kamiokande \cite{SK1,SK2}
and SNO \cite{SNO} experiments are studied in sects. 7, 8.
In sect. 9  we summarize our main results and
also comment on possible  modifications of the $GU$-scenario
which would allow one  to explain other neutrino data.

\vskip 0.3truecm

%%%%%%%%%%%%%%%%%%%%%%%%%%%%%%%%%%%%%%%%%%%%%%%%%%%%%%%%%%%%%%%%%%%%%%%%%%%

\leftline{\bf 2. $\nu_e~ -~ \nu_{\tau}$ conversion
of solar neutrinos
and the scale of Grand Unification}

\vskip 0.2truecm

\indent
Let us  focus on the mass of the third neutrino $m_{3}$
which determines $\Delta m^2_{13}$ responsible for the
$\nu_e - \nu_{\tau}$ conversion.
We perform first a calculation of $m_{3}$
neglecting $\nu_{\tau}$ mixing with  lighter generations.
The procedure adopted is as follows.
We use  (\ref{mSeeSaw}), (\ref{mDup})
 and the two--loop renormalization group equations
(RGE) with the particle content of
the Minimal Supersymmetric  Standard Model.
We fix   the QCD gauge coupling  $\alpha_3(M_Z) \equiv
 g_3^2(M_Z)/4 \pi = 0.118$ at the $Z^0$ boson mass scale, $M_Z$,
as an input and set the effective supersymmetry
threshold at $M_Z$.

The renormalized neutrino mass at a scale $\mu < \Lambda_{GU}$ with the
boundary condition (\ref{mDup}) can be written as
\begin{equation}
m_{3}(\mu) = {m_t^2(\mu) \over M_{03} }
\left[{\kappa(\mu) \over \kappa(\Lambda_{GU})} \right]
\left[ {h_t^2(\Lambda_{GU}) \over h_t^2(\mu)} \right]~,
\label{nurge}
\end{equation}
where $\kappa$ is the coefficient of the effective dimension-5 neutrino
mass operator, ${\cal L}_{eff} = \kappa_{ij} L_i L_j H_u H_u$,
$M_{03}$ is the mass of the RH neutrino in the absence of mixing between
generations.
The RGE for $\kappa$
(which has been worked out only to one loop)
is given by \cite{blp}
\begin{equation}
{d \kappa \over dt} =
{1 \over 16 \pi^2} \left[6h_t^2+2h_\tau^2-{6 \over 5} g_1^2
-6 g_2^2 \right]~. \nonumber
\label{kappa}
\end{equation}
Here $g_1$ and $g_2$ are the $U(1)$ and $SU(2)$ gauge  couplings,
and $h_\tau$ is the Yukawa couplings of the tau lepton.
The  one--loop RGE for  the top Yukawa coupling $h_t$ is
\begin{equation}
{dh_t^2 \over dt} =
{h_t^2  \over 8 \pi^2} \left[6h_t^2+h_b^2-{13 \over 15} g_1^2
-3 g_2^2 -{16 \over 3} g_3^2 \right]~,  \nonumber
\label{top}
\end{equation}
where $h_b$ is the bottom Yukawa coupling.

In our numerical calculations of $m_3$ we kept the full two--loop RGE
equations for $h_t$, $h_b$, $h_\tau$. The results are summarized in
Table I, where
the values  of $m_{3}$ are shown for different values of $\tan\beta
\equiv {v_u / v_d}$ and  three possible values of $m_t$; the Majorana mass
is fixed to be $M_{03} = 10^{16}~{\rm GeV}$.

As follows from Table I
the $\nu_\tau$--mass is insensitive to the value of
tan$\beta$ except for tan$\beta \lesssim 2$.
This feature can be understood from the one-loop semi--analytic
expression for $m_{3}$. Indeed,
the neutrino mass at a lower scale $\mu$ can be  written as
\begin{equation}
m_{3}(\mu) = {m_t^2 (\mu) \over M_{03} }
\left({\alpha_1(\mu) \over \alpha_G }\right)^{4/99}
\left({\alpha_3(\mu) \over \alpha_G} \right)^{-16/9} \times
\nonumber \\
exp \left[{1 \over 8 \pi^2}
\int_{{\rm ln}\mu} ^{{\rm ln}\Lambda_{GU}}dt
\left(3h_t^2 + h_b^2 - h_\tau^2 \right)
\right]~,
\label{numass}
\end{equation}
where $\alpha_G \simeq 1/25$ is the unified gauge coupling constant
and $\alpha_1 \equiv g^2_1/ 4\pi$.
The neutrino mass depends
on  $\tan \beta$  via  the Yukawa coupling
$h_t = m_t \sqrt{1+{\rm tan}^2\beta}/(v {\rm tan}\beta)$.
Obviously, for  tan$\beta \ge 2$, the coupling  $h_t$ is essentially
independent of tan$\beta$.

Using the results of Table I and the mass of the light neutrino
suggested by the solar neutrino data (\ref{mMSW}) we can estimate the
RH-neutrino mass $M_{03}$.
For $m_t = 175$~GeV  and tan$\beta = 1.7$
we find
$M_{03} = (0.5 - 0.7) \Lambda_{GU}$.
For tan$\beta \lesssim 2$  the top Yukawa coupling
is near its fixed point value.  Such a low value
may be preferable from the point of view of successful $b-\tau$ unification.
On the other hand, for tan$\beta \ge 2$, we get
$M_{03} = (0.18 - 0.25) \Lambda_{GU}$.
Thus for the whole range of tan$\beta$ the $\Delta m^2$
required for solution of the $\nu_{\odot}$-problem corresponds to
$M_{03}$ in the interval
\begin{equation}
M_{03} \lesssim \displaystyle (0.2~-~0.7)~ \Lambda_{GU}.
\label{mismatch}
\end{equation}

One remark is in order.
If  $M_{03}$ is substantially  below $\Lambda_{GU}$,
one should take into account  the
effect of the Dirac neutrino Yukawa coupling
$h_{3D}$ on the evolution of
$h_t$, $h_b$, $h_\tau$ and $h_{3D}$ itself in the interval
between $M_{03}$ and  $\Lambda_{GU}$, since in this interval
loops involving $\nu_{3R}$ will contribute \cite{sv}.
Although the momentum range for the running is rather small the effect
can be substantial since
both $h_t$ and $h_{3D}$ are  large
(near the fixed point value of $h_t$).
The boundary condition at $M_{03}$
on the Dirac Yukawa coupling is modified to
\begin{eqnarray}
{h_{3D}^2(M_{03}) \over h_t^2(M_{03})}  = {h_{3D}^2( \Lambda_{GU})
\over
h_t^2( \Lambda_{GU}) } \left({\alpha_1(M_{03}) \over \alpha_G}\right)^{4/99}
\left({\alpha_3(M_{03}) \over \alpha_G} \right)^{-16/9} \times  \nonumber
\\
exp \left[{1 \over 8 \pi^2} \int_{{\rm ln} \Lambda_{GU}}^{{\rm ln}M_{03}} dt
\left(3h_{3D}^2-3h_t^2 +h_\tau^2-h_b^2\right)\right]~. \nonumber
%\label{threshold}
\end{eqnarray}
In the relatively  small momentum range
between the scales $M_{03}$ and $ \Lambda_{GU}$
the exponential factor above is
almost unity because $h_{3D} = h_t$ and $h_b = h_\tau$
at $ \Lambda_{GU}$.
A difference between $h_t$ and $h_{3D}$  develops because of
the difference in the gauge contribution.
The effect of the gauge couplings is to diminish the Dirac mass
$m_{3D}$, and consequently, the mass $m_3$ at low scale.
%and 1/5, we find $h_{\nu_\tau}^2(M_{03})/h_t^2(M_{03}) = 0.92$
%and  0.94 correspondingly.  Now the boundary
%condition on the see--saw formula must be applied at $M_{03}$ rather than
%$ \Lambda_{GU}$.  This effect alone changes the predicted value of $m_{3}$
%from the
%values in the Table I  to  $(2.3,~ 1.0,~ 0.89,~ 0.91,~ 1.04)\cdot
%10^{-3}~eV$ for
%
Indeed, for $M_{03}/\Lambda_{GU} = 1/10$,   $m_t = 175$~GeV  and  the
set of values tan$\beta = (1.7,~ 3,~ 10,~ 30,~ 60)$  we find
the neutrino mass
$m_{3} = (2.15,~ 0.95,~ 0.82,~ 0.84,~0.96) \cdot 10^{-3} (10^{16}
{\rm GeV}/M_{3})$ eV.
Comparing these values with the numbers  in  Table I we conclude that
the threshold effect related to $M_{03}$
decreases the predicted value of $m_{3}$ by 5 - 25 \%.
Equivalently, the predicted value of $M_{03}$ for
a fixed $m_{3}$ decreases  by as much as 25\% for tan$\beta =
1.7$ (fixed
point value), and by   6\%  for moderate and  large 
tan$\beta$ ($M_{03}/\Lambda_{GU} = 1/10$).
So, the  values of $M_{03}$ are not changed
significantly from those in (\ref{mismatch}).

%$$m_{\nu_\tau}~\simeq~(0.48~-~0.49)\cdot 10^{-3}\cdot \left({m_t \over
%175~
%\rm {GeV}} \right)^2 \cdot
% \left({10^{16}~ {\rm {GeV}} \over {M_R}} \right)~\rm {eV}.$$

Let us now take into account mixing of the third generation with
the light generations of fermions.
For simplicity we will consider mixing between two
(say second and third) generations.
Suppose  $M_{02}$ is
the Majorana mass of the RH neutrino component
which leads  via the see-saw mechanism to a certain mass
$m_2$ in the absence of mixing between generations.
For fixed  values of
$m_{2}$ and $m_{3}$ as well as the Dirac masses $m_{2D}$, $m_{3D}$,
the mixing in the Majorana mass matrix of the RH neutrinos,
$\hat M$, will change values of the corresponding
RH neutrino masses:
$M_{2} \neq M_{02}$ and
$M_{3} \neq M_{03}$, where
$M_{2}$ and  $M_{3}$ are the masses in presence of mixing.
It can be shown, however,
that the product of masses does not depend on the mixing,
that is, $M_{2} \cdot M_{3} = M_{02}\cdot M_{03}$
(up to  small renormalization group corrections) \cite{smirnovint}. From
this equality we get a relation between hierarchies
of masses in the presence of mixing,
$\epsilon \equiv  M_{2}/M_{3}$,  and without mixing,
$\epsilon_0 \equiv M_{02}/M_{03}$:
\begin{equation}
\epsilon = \epsilon_0 \left(\frac{M_{03}}{M_3}\right)^2~.
\label{hierar}
\end{equation}
The change $M_{03} \rightarrow M_3$ is related to
the mixing angle $\theta_M$ in $\hat M$  in the following way
\cite{smirnovint}:
\begin{equation}
\sin^2 \theta_{M}~\approx~
\epsilon \left[\frac{M_3}{M_{03}} - 1\right]~.
\label{anglem}
\end{equation}

The mixing in $\hat M$ can raise  the mass
$M_{3}$ to the $GU$ - scale:
$M_{03} \rightarrow  M_{3} \sim \Lambda_{GU}$,
for fixed  masses of the light neutrinos
and for fixed Dirac masses. This  implies that
the second mass, $M_2$,  decreases by  factor
${M_{03}}/{M_3}$,
and according to (\ref{hierar}) the hierarchy of masses
is enhanced  by a factor $({M_{03}}/{M_3})^2$.
If $M_{3}$ increases, {\it e.g.},  by a factor 3,
the hierarchy of masses is enhanced by one
order of magnitude.
Similar results can be obtained for  mixing between the
third and the first generations.

~From (8) we see that the mass of $\nu_3$ in the absence of mixing,
$M_{03}$,  is (1.5 - 5) times smaller than $\Lambda_{GU}$.
As follows from (\ref{anglem}) small  mixing between  generations
$\theta_M \sim O(\sqrt {\epsilon})$
is enough to get the equality  $M_3 = \Lambda_{GU}$.
Another consequence of the increase in $M_{3}$ and the
strengthening of the hierarchy is the enhancement
of mixing of the light
neutrinos which will be discussed in the sect. 3.

The proximity of $\Lambda_{GU}$ and  $M_{3}$ suggested by the solar
neutrino data (or even possible coincidence
of these two values) can be considered as a hint
for Grand Unification.
Let us comment on  possible relations between these two  scales.

Since the neutrino mass term $M  \nu_R^T \nu_R$ is a
singlet of the standard model symmetry group, there is no {\it direct}
relation between the neutrino mass $M$ and the scale of the
gauge coupling unification.
Some additional assumptions are needed to connect
$M_{3}$ and $\Lambda_{GU}$  and, in general,
one should not expect a coincidence of these scales.
Possible relations between $M_{03}$ and $\Lambda_{GU}$
depend on the mechanism of $\nu_R$  mass generation.
Two simple mechanisms are worth mentioning:

(i) The Majorana mass $M_{3}$ is generated directly by
the Yukawa couplings with a Higgs multiplet $\Phi$
\begin{equation}
h_{3M} \nu_R^T \nu_R \Phi ~,
\label{yukawa}
\end{equation}
so that $M_{3} = h_{3M}  \langle \Phi \rangle_0$.
A relation between  the interaction (\ref{yukawa})
and  physics of Grand Unification
can arise in  $SO_{10}$ models, where
$\Phi$ is the $\overline{\bf 126}$-plet.
If $\Phi$ is responsible for the symmetry breaking
$SO_{10} \rightarrow SU_5$ at the $GU$-scale,  then
$\langle \Phi \rangle_0 \sim \Lambda_{GU}$.
Still  $h_{3M}$ should be fixed.
Note that $h_t(\Lambda_{GU}) \simeq (0.4-1.5)$ for a wide range of
tan$\beta$. Therefore,
to get  $M_{03}$ in  the region (\ref{mismatch})   one should take
$h_{3M} \sim (0.13 - 1.4) \cdot h_t$ in the absence of mixing.
This allows for the possibility, especially if
there is some mixing in the mass matrix $\hat M$, that
all the Yukawa couplings of fermions  from  the
third generation are the same, in particular,   $h_{3M} \sim h_t$.

(ii) The mass  $M_3$ can be generated  by the
effective operator
\begin{equation}
\frac{h_{R}^2}{M_S} \nu_R^T \nu_R \Phi \Phi~,
\label{effect}
\end{equation}
where $M_S$ is some mass at or above the $GU$-scale.
In this case  $M_3 \simeq h_R^2 {\langle \Phi \rangle}^2_0/M_S$.
In  $SO_{10}$  models $\Phi$ is $\overline{{\bf 16_H}}$-plet,
and the operator (\ref{effect})
is generated by terms in the superpotential
\begin{equation}
 h_R{\overline {\bf 16}}_H {\bf 16} S~+~M_S S S~
\label{r1616S}
\end{equation}
which mix $\nu_R$ with an $SO_{10}$-singlet or an adjoint, $S$,
having the Majorana mass $M_S$.  The
$\overline {{\bf 16}}_H$-plet  breaks
$SO_{10}$ to  $SU_5$ and it is possible  to identify its VEV
with $\Lambda_{GU}$. For
$M_S \sim (1 - 2) \cdot \Lambda_{GU}$ one can take
$h_{3M} \sim  h_t$ in agreement with equality of
all Yukawa couplings in the third generation.\\

%%%%%%%%%%%%%%%%%%%%%%%%%%%%%%%%%%%%%%%%%%%%%%%%%%%%%%%%%%%%%%%%%%%%%%

\vskip 0.3truecm
\leftline{\bf 3. \nue --  \nutau mixing}
\vskip 0.2truecm

In the two
neutrino case the mixing angle needed to solve the $\nu_{\odot}$-problem
should be  in the range determined by \cite{krastev}:
\begin{equation}
\sin^2 2\theta~=~(3~-~10) \cdot 10^{-3}.
\label{sinMSW}
\end{equation}
As we will see in sect. 4, the effect of mixing between \nue and \numu
enlarges the region of solutions (\ref{sinMSW}) to:
\begin{equation}
\sin^2 2\theta_{e\tau} ~=~ (0. 2~-~10) \cdot 10^{-3}.
\label{sinMSWVO}
\end{equation}
Let us first compare (\ref{sinMSWVO})  with the corresponding
mixing in the quark sector.  Taking $~\theta_{e\tau}
\sim V_{td}\sim(4~-~11) \cdot 10^{-3}~$, where ${V_{td}}$ is the
element of the CKM quark mixing matrix, we
get $~\sin^2 2\theta_{e\tau} ~=~(0.06~-~0.5) \cdot 10^{-3}$.
These  values  cover the lower part of
the range  (\ref{sinMSWVO}).
Thus, the $\nu_e  -  \nu_{\tau}$ mixing similar to the quark
mixing is enough to solve the $\nu_{\odot}$- problem
provided the  $\nu_e - \nu_{\mu}$ mixing is sufficiently large:
$\sin^2 2 \theta >  0.3$  (see sect. 6).
For $\sin^2 2 \theta_{e\mu} < 0.3$ the values
$\theta_{e\tau} \sim V_{td}$
are too small. In this connection
let us consider the possibility of enhancing
the mixing, so that $\theta_{e\tau} \sim (2 - 3)~ V_{td}$.

In view of the violation of  quark-lepton symmetry
among lighter generations
there is no reason to expect exact equality of
$\theta_{e\tau}$ and $V_{td}$.
A difference in quark and lepton mixing can come  both
from difference in the Dirac mass matrices: 
$m_D \neq m_q$ (violation of the quark-lepton symmetry)
and from the Majorana mass matrix (see-saw enhancement of
lepton mixing \cite{enhan}). Let us consider these two possibilities in
order.

1) The quark-lepton symmetry
 for light generations can be broken by
the Yukawa couplings
with additional Higgs multiplets.
The contributions from  these couplings
to masses correct the bad asymptotic mass relations
$m_d(\Lambda_{GU}) = m_e(\Lambda_{GU})$ and
$m_s(\Lambda_{GU}) = m_\mu(\Lambda_{GU})$.
Such corrections tend to enhance the mixing in the lepton sector.
An explicit example has been given in the context of
$SO_{10}$  \cite{babumohap}.
There a minimal set of Higgs multiplets was introduced that
couples to fermions:  namely,  a single ${\bf 10}$-plet
and a single ${\bf \overline{126}}$-plet.
The Standard Model singlet from  the
${\bf \overline{126}}$-plet  generates
the Majorana mass term for the RH neutrinos.
The same ${\bf \overline{126}}$-plet  also contains a pair of
SM doublets which receive vacuum expectation values,
contributing to the quark and lepton masses.
As a consequence, the structure of the Majorana mass matrix
gets related to the quark and lepton mass matrices.
The  Yukawa couplings of
${\bf \overline{126}}$-plet
correct the bad asymptotic mass relations
$m_d(\Lambda_{GU}) = m_e(\Lambda_{GU})$ and $m_s(\Lambda_{GU}) =
m_\mu(\Lambda_{GU})$.
This led to the prediction
$\theta_{e \tau} \sim 3 V_{td}$, where
the factor 3 is the same Clebsch-Gordon
coefficient that corrects the
mass relations for light generations \cite{gj}.
In this case one gets $\sin^2 2\theta_{e\tau} = (0.5 - 4) \cdot 10^{-3}$
 -- within the range (\ref{sinMSWVO}).

2).  An  enhancement of lepton mixing can follow from
mixing in  the Majorana mass matrix of the RH neutrinos
 \cite{enhan}. Indeed, the total lepton mixing angle
(for two generation case) can be written as
\begin{equation}
\theta~=~\theta_D~+~\theta_{s}~,
\label{totalmixing}
\end{equation}
where  $\theta_D$ follows from  the Dirac mass
matrices of the neutrinos and charge leptons
($\theta~=~\theta_D$, if $\hat{M} = \hat{I}$)   and
$\theta_{s}$ specifies the effect of the see-saw mechanism itself
\cite{enhan}. Let us consider  mixing between
the first and the third generation.
The see-saw angle $\theta_s$ can be expressed in terms of the mass
hierarchies $\epsilon_D \equiv m_{1D}/m_{3D}$,
$\epsilon \equiv M_1/M_3$ and  $\epsilon_0 \equiv M_{01}/M_{03}$
\cite{smirnovint}:
\begin{equation}
\sin^2 \theta_{s}~\approx~
\frac{\epsilon_D^2}{\epsilon}
\left[ \sqrt{\frac{\epsilon_0}{\epsilon}} - 1
\right] =
\frac{\epsilon_D^2}{\epsilon}
\left[\frac{M_3}{M_{03}} - 1\right]~.
\label{sees}
\end{equation}
For  {\it linear} mass hierarchy
of the RH neutrinos, $\epsilon_0 \sim \epsilon_D$, we get
\begin{equation}
\sin^2 \theta_{s}~\approx~
\epsilon_D \frac{M_3}{M_{03}}~.
\label{sees1}
\end{equation}
Therefore  $\epsilon_D = 10^{-5}$ and
$M_3/M_{03} \sim 2 $ lead to
$\theta_s \sim 3 \cdot 10^{-3}$,
which is comparable with $V_{td}$. If the hierarchy of
masses is strengthened  by one  order of magnitude
($\epsilon = 0.1 \epsilon_0$),
then the angle  $\theta_s$ becomes  $10^{-2}$.
In this case $\theta_s$  gives the main contribution to the lepton
mixing,  and $\sin^2 2\theta_{e\tau}$ is of the order  $10^{-3}$.

According to (\ref{anglem}) for the linear hierarchy one gets
$\theta_M \sim \sqrt{\epsilon} \sim \sqrt{\epsilon_D}$. That is,
the mixing in $\hat M$ similar to mixing in the Dirac mass matrices can lead
to enhancement of the lepton mixing up to the value implied by the
$\nu_{\odot}$-problem. Moreover, as we discussed in sect. 2,  this
enhancement can be related to increase of
$M_3$ to $\Lambda_{GU}$.

There is an alternative mechanism for generating enhanced lepton
mixing.  Mixings in the lepton sector can differ substantially
from those in the quark sector as a consequence of the
nature of
Grand Unification symmetry breaking.  The stronger
hierarchy observed in the up--quark masses relative
to the hierarchies in the down
quarks and charged lepton masses gives some indication towards
such a possibility. One  example was
suggested in \cite{bb}.
The main idea  is that in $SU_5$ (or
in the $SU_5$ decomposition
of $SO_{10}$) $u_i$ and $u_i^c$ quarks lie in the same ${\bf 10}_i$
representation, while the $d_i$ and $d_i^c$ (and similarly
leptons $l^c_i$ and $l_i$ )
are  in separate representations:
${\bf 10}_i$ and ${\bf \overline{5}}_i$
respectively.  If a hierarchy factor
$\epsilon_D$ is associated with the ${\bf 10}$-plets, but
not with the ${\bf \overline{5}}$-plets,
the down quark and charged lepton
masses  (${\bf \overline{5}}$
$\cdot {\bf 10}$ )
will scale as $\epsilon_D$,  whereas the up-quark masses
(${\bf 10}$ $\cdot {\bf 10}$) will scale as $\epsilon_D^2$.
The left--handed CKM angles will be
small, as they scale as $\epsilon_D$.
The right--handed CKM mixing angles for quarks
(unphysical in the Standard Model) will be of order one,
since the ${\bf \overline{5}_i}$ have no hierarchy factor.
Since  the left--handed  lepton doublets are
in ${\bf \overline{5}}_i$, their
mixing angles will be large.
The specific model constructed
in Ref. \cite{bb} leads to $\theta_{e\tau} \simeq 0.032$ which
is within the range (\ref{sinMSWVO}).

%%%%%%%%%%%%%%%%%%%%%%%%%%%%%%%%%%%%%%%%%%%%%%%%%%%%%%%%%%%%%%%%%%%%

\vskip 0.3truecm
\leftline{\bf 4. Parameters of \nue -- \numu  system}
\vskip 0.2truecm

If the mixing angles
in the Majorana matrix $\hat {M}$ are small, the mass
of the second neutrino can be estimated as
\begin{equation}
{\displaystyle m_2 ~\approx~ -{(h_c v_u)^2 \over {M_{02}}}}~,
\label{mass2}
\end{equation}
where $h_c$ is the Yukawa coupling of the charm quark.
For ${M_{02}} \approx {M_{03}}\approx \Lambda_{GU}$
we find $~m_2 \leq 10^{-7}$ eV
and $~\Delta m^2_{12} \approx 10^{-14}$ eV$^2~$  which is far below
the sensitivity of the $\nu_{\odot}$-data.
Let us assume
a  {\it linear} mass hierarchy in the RH neutrino mass matrix,
according to which
the eigenvalues of $\hat{M}$ have the same
hierarchy as the eigenvalues of the Dirac mass matrix
$m_{D}$:  $M_{i} \propto m_{iD}$, or $\epsilon \sim \epsilon_D$.
In this case $M_{02} \sim 10^{14}$ GeV, and
for the light neutrino we get from the see-saw
formula (\ref{mSeeSaw})
$m_2 \approx (0.3 ~-~3) \cdot 10^{-5}$ eV. Consequently,
\begin{equation}
\Delta m^2_{12} ~\approx~ (10^{-11}~-~10^{-9}~)~\rm {eV}^2~,
\label{delmvac}
\end{equation}
which is in the range where vacuum oscillations on the way
from the Sun to the Earth are important.

Let us comment on the possible origin of  the linear mass hierarchy of
$\hat{M}$.
If in  $SO_{10}$ models   the same
$\overline{\bf 126}$-plet contributes
both to the RH neutrino masses and
to the charged fermion masses, the hierarchy is
naturally linear  \cite{dhr}.

The linear hierarchy may be due to family symmetry.
Let us consider a $U(1)$ symmetry whose
breaking is characterized by a single small
parameter $\lambda$ (which may be the VEV of
a singlet Higgs scalar divided by the Planck mass)
having the $U(1)$-charge $- 1$.
We assume that the fermionic
${\bf 16}_i$-plets  of $SO(10)$ carry
family $U(1)$   numbers $q_i$.
Then  as a consequence of the $U(1)$ invariance, the mass term
${\bf 16}_i{\bf 16}_j$ will be suppressed by a
factor $\lambda^{q_i+ q_j}$.
Since  the Majorana mass terms
((\ref{yukawa}) or (\ref{effect})) have the same flavor structure
as the Dirac mass terms
they will have similar  hierarchy in the eigenvalues
provided that the only source of $U(1)$--breaking is $\lambda$.

Note that in the case of linear mass hierarchy the
lightest RH neutrino has the mass
$(10^{-6} - 10^{-5})~\Lambda_{GU} \sim (10^{10} - 10^{11})$ GeV
which is in the correct range to explain  baryogenesis via
leptogenesis \cite{bp}.

The $\nu_e-\nu_\mu$ mixing angle is expected to  be of
the order the  Cabibbo angle, $\theta_c$. It can be
larger than $\theta_c$ due to violation
of exact quark--lepton symmetry in the light generations.
Furthermore, an  additional enhancement may follow from the
see-saw mechanism itself, as was discussed in sect. 3.
Therefore   $\theta_{e\mu}\sim (1 - 2)~\theta_{c}$,
and consequently, $\sin^2 2\theta_{e\mu} \sim 0.2 - 0.7$  are  quite
plausible  without any additional assumptions.
This mixing can lead to observable effects in the solar neutrinos.

In the $GU$-scenario the masses of $\nu_1$ and $\nu_2$ are so small
even in the case of the linear mass hierarchy of the RH components
that  possible contributions from the Planck scale physics
can be important. In general, an expression  for neutrino masses
is the sum
\begin{equation}
$$ m~\approx~m_{s}~+~m_{nr}~, $$
\label{Tm}
\end{equation}
where $m_{s}$ is  the see-saw mass and $m_{nr}$ is
the contribution from possible effective non-renormalizable
interactions associated to
the Planck scale physics \cite{weinberg}:
\begin{equation}
$$ {\displaystyle {\alpha_{ij} \over {M_P}}L_jL_iH_uH_u}~. $$
\label{nonnor}
\end{equation}
Here $\alpha_{ij}$ are constants expected to be of  order 1, $L_i$ are the
leptonic doublets  and $M_P$ is the Planck
mass.
The interaction (\ref{nonnor}) may follow immediately from
string compactification
or from  renormalizable interactions via the exchange of
particles with  mass $\sim~M_P$.

The interaction (\ref{nonnor}) gives the mass
${\displaystyle m_{nr}~\sim~ {\alpha_{ij} \over
{\it {M_p}}}} v_u^2 ~\sim~ 10^{-5}$ eV.
%If all $\alpha_{ij}$ are of the same order, then
The contribution $m_{nr}$ to the mass of the
third neutrino can be neglected,  whereas $m_{nr}$
is important for  $m_2$ and it
dominates in $m_1$. Again, $\Delta
m^2_{12} \sim m_{nr}^2
\sim 10^{-10}$ eV$^2$ is in the range (\ref{delmvac}).
Also the mixing is modified.
The interactions
(\ref{nonnor}) determine the
mixing between the first and second neutrinos.
If  $\alpha_{ij}
\sim O(1)$, this  mixing angle can be large: $\sin^2
2\theta_{e\mu}~\sim~O(1)$.
The mixing of light generations with the third
generation remains small: {\it e.g.},
 $\theta_{e\tau}  \sim m_{nr}/m_{3} \sim 0.01$.
Nevertheless this contribution is
 comparable to $V_{td}$.
Therefore the total $\nu_e - \nu_{\tau}$ mixing can be enhanced by
interactions (\ref{nonnor}), so that
$\sin^2 2\theta_{e\tau}$ will be in the interval
(\ref{sinMSW}).

%%%%%%%%%%%%%%%%%%%%%%%%%%%%%%%%%%%%%%%%%%%%%%%%%%%%%%%%%%%%%%%%%%%%

\vskip 0.3truecm
\leftline{\bf 5. The interplay of resonance conversion and vacuum
oscillations}
\vskip 0.2truecm

As we have established in sects. 3, 4  the $GU$- scenario
with linear mass hierarchy of the RH neutrino masses or/and
with additional effects from  the non-renormalizable
Planck scale interactions can naturally provide
the pattern of the neutrino masses and mixing with
$m_3~\sim~ (2~-~3)\cdot 10^{-3}$ eV,
$m_2~\sim~(0.3~-~3)\cdot 10^{-5}$ eV,
$m_1 < m_2$, $\sin^2 2\theta_{e\tau} = (0.2 - 3)\cdot 10^{-3}$
and
$\sin^2 2\theta_{e\mu} \sim  0.2 - 0.7$.
For these values of parameters both
the $\nu_e \rightarrow \nu_{\tau}$
resonance flavor conversion
and $\nu_e - \nu_\mu$  vacuum oscillations on the way from the
Sun to the Earth become important.  Moreover,
non-trivial interplay of these two effects takes place.
We will call such a possibility the {\it hybrid}
(resonance conversion plus vacuum oscillation)  solution
of the solar neutrino problem.

Due to the mass hierarchy and the smallness
of $\theta_{e\tau}$ the dynamics of the three neutrino system   is
reduced to a
two neutrino task. Indeed, the electron neutrino state can be written in
terms of the mass eigenstates $\nu_i~(i=1,2,3)$ as
\begin{equation}
\nu_e~=~\cos \theta_{e\tau} \cdot \tilde {\nu}~+~\sin \theta_{e\tau}
\cdot \nu_3~,
\label{nuef}
\end{equation}
where
$$
\tilde {\nu} \equiv
\cos \theta_{e\mu} \nu_1+\sin \theta_{e\mu} \nu_2~.
$$
Inside the Sun due to the smallness of
$\Delta m^2_{12}$, the system $\nu_1~-~\nu_2$ is ``frozen''.
The evolution of the whole system consists of
the \nue resonance conversion  to the state
$\nu'=\cos \theta_{e\tau} \nu_3 - \sin \theta_{e\tau} \tilde {\nu}$
(orthogonal to \nue).
On the way from the surface of the Sun to the Earth the state
\nuthree decouples from the system:
large mass difference $\Delta m^2_{13}$ leads to averaged
oscillation effect or/and to loss of coherence.
On the way to the Earth the $\nu_e - \nu_{\mu}$
vacuum oscillations  occur due to mass splitting
$\Delta m^2_{12}$  between \nuone and \nutwo.
Taking this into account it is easy to write the
\nue survival probability \cite{AS91}:
\begin{equation}
P~=~P_V(\Delta m^2_{12},~\theta_{e\mu}) \cdot {P}_R(\Delta m^2_{13},
~\theta_{e\tau})~ +
~O(\sin^2 \theta_{e\tau}).
\label{Pf}
\end{equation}
Here $P_V(\Delta m^2_{12}, ~\theta_{e\mu})$
is the 2$\nu$ - vacuum oscillation
probability characterized by parameters
$\Delta m^2_{12},~\theta_{e\mu}$, and
${P}_R$ is the 2$\nu$ - averaged
survival probability of the resonance conversion
characterized by $\Delta m^2_{13}$ and $\theta_{e\tau}$
($P_R$ is averaged over the production region).
For $\sin^2 2\theta_{e\tau} \leq 3 \cdot 10 ^{-3}$
the $O(\sin^2 \theta_{e\tau})$ -
corrections in (\ref{Pf}) can be safely neglected and the
total probability is  factorized.
The exact formula has been derived in \cite{QLSP}.
In fig. 1 we show typical  dependence of the  survival probability
on the neutrino energy.

The survival probability (\ref{Pf}) satisfies the following
inequalities:
\begin{equation}
(1~-~\sin^2 2\theta_{e\mu}) \cdot {P}_R~\leq~P~\leq~{P}_R~.
\label{Pupb}
\end{equation}
That is,  $P$ is an oscillatory function of the neutrino
energy inscribed in the band  (\ref{Pupb}) (fig. 1).
The width of the band equals
$\Delta P = \sin^2 2\theta_{e\mu} \cdot {P}_R$.

If vacuum oscillations are averaged out, we get from (\ref{Pf})
\begin{equation}
P ~=~ (1~-~{1 \over 2} \sin^2 2\theta_{e\mu}) \cdot {P}_R~.
\label{Pf2}
\end{equation}

The properties (\ref{Pupb}, \ref{Pf2})
allow one to derive immediately several consequences of the
interplay of the vacuum oscillations and resonance conversion.

\begin{itemize}

\item
The solar  $\nu_e$ are  transformed both to $\nu_{\mu}$
and  $\nu_{\tau}$:
the corresponding transition probabilities equal
$P(\nu_e \rightarrow \nu_{\tau}) \approx  1 - P_R $ and
$P(\nu_e \rightarrow \nu_{\mu}) \approx   P_R - P$,
where $P$ and $P_R$ are defined in (\ref{Pf}).

\item
According to (\ref{Pupb}), vacuum oscillations lead to additional
suppression of the \nue-flux in comparison with the  pure resonance conversion.
As a consequence, new regions of neutrino parameters appear
in which one can  describe the $\nu_{\odot}$-data. In particular,
$\sin^2 2\theta_{e\tau} < 10^{-3}$ are allowed now (sect. 6, fig. 2).

\item
For  $\Delta m_{12}^2 \leq 10^{-9}$ eV$^2$  the vacuum
oscillations are not averaged and lead to additional
oscillatory distortion of the $^8 B$-
or/and $pp$-neutrino spectra (sect. 7).

\item
One expects additional seasonal variations of neutrino flux due to the
dependence of the vacuum oscillation probability on the distance
from the Sun to the Earth
(sect. 8).

\item
For large $\Delta m_{12}^2$
the $pp$- neutrino
flux is  suppressed by the averaged vacuum oscillation effect.
For small  $\Delta m_{12}^2 <  10^{-11}$ eV$^2$
the suppression depends on the neutrino energy,
thus leading to distortion of
the $pp$-neutrino energy  spectrum.

\end{itemize}

Thus, a detailed study of the energy spectra of the $^8B$-
neutrinos,  and (in future)  of  $pp$- neutrinos
as well as measurements  of seasonal
variations of the signals will allow one to test the $GU$-scenario.

%%%%%%%%%%%%%%%%%%%%%%%%%%%%%%%%%%%%%%%%%%%%%%%%%%%%%%%%%%%%%%%%%%%%%%%%%%
\vskip 0.3truecm
\leftline{\bf 6. New regions of parameters}
\vskip 0.2truecm

The interplay of the resonance conversion and vacuum oscillations opens
new possibilities in description of  the $\nu_{\odot}$-data \cite{AS91}.
Some of these  possibilities were discussed in \cite{QLSP}.
In particular, for  $~\sin^2 2\theta_{e\tau} \sim  10^{-3}$   and
$\Delta m^2_{13} \sim 10^{-4}$ eV$^2$
the high energy part of the
boron neutrino spectrum lies
in the nonadiabatic edge of the two
neutrino suppression pit (due to the resonance conversion).
For these values of parameters  the resonance conversion
does not change neutrino fluxes at low and intermediate energies.
The suppression of the $^7 Be$-neutrino flux implied by the
$\nu_{\odot}$-data can be due to vacuum oscillations,
if {\it e.g.}, the first high energy dip of the
oscillation probability  is at $E$ $\sim$ 1 MeV.
For this to occur, $\Delta m^2_{12}$
should be in the range (7 - 9)$\cdot 10^{-12}$ eV$^2$.
Thus,
the vacuum oscillation suppression pit
is at low energies, and the resonance conversion pit is at high energies.
Note that  $\Delta m^2_{13} \sim 10^{-4}$ eV$^2$
correspond to   values of $M_{3}$
substantially below the $GU$-scale:
$M_{3} \sim 5\cdot 10^{14}$ GeV.

In contrast, for the $GU$-scenario with $\Delta m^2_{13}=
(4~- ~10 )\cdot10^{-6}$ eV$^2$
the $^7 Be$ -flux can be suppressed by the resonance conversion. Since
$~\sin^2 2\theta_{e\tau} \lesssim 10^{- 3}$, the
suppression pit is narrow
and  the high energy (observable) part of the boron
neutrino spectrum is suppressed insufficiently.
An additional  suppression can be  due to vacuum oscillations,
especially if the observable  part of the boron neutrino flux  is situated
in the first
high energy dip  of the oscillatory curve (see fig. 1).
Thus we get the configuration with the resonance conversion pit
at low  energies and the vacuum oscillation pit at high energies.
The region of parameters, where
this configuration gives a good fit of the data
is shown in fig. 2. (In the $\chi^2$-analysis
we considered the Super-Kamiokande and Kamiokande
as  separate experiments.
Thus,  there is one degree of freedom for the hybrid solution
and three degrees of freedom for
2$\nu$ solution of the solar neutrino problem.)
As follows from fig. 2  values of mixing angles as small as
$\sin^{2} 2\theta_{e \tau} = 2 \cdot 10^{-4}$ are now allowed.

Further diminishing of $\theta_{e\tau}$ implies an increase of
$\theta_{e\mu}$ in such a way that
the hybrid solution converges to pure vacuum oscillation solution.
Furthermore, it is impossible to  diminish $\Delta m^2_{13}$
and therefore to increase $M_3$
in the hybrid solution.
The  minimal value, $\Delta m^2_{13} \approx 4 \cdot 10^{-6}$ eV$^2$,
is defined by the
condition that $^7 Be$-neutrino line is at the adiabatic edge of the
resonance conversion pit.

For the $pp$-neutrinos one gets the averaged oscillation effect:
$P \approx 1 - 0.5 \sin^2 2\theta_{e \mu} \sim 0.7 - 0.9 $.
Therefore, the Germanium production rate in
the Gallium experiments is typically at the lower border of the allowed
region ($\sim 60$ SNU's). This suppression is  weaker
than the one in the  pure vacuum oscillation solution.

Since the mixing angle responsible for the resonance
conversion is very small, $\sin^2 2\theta_{e \tau} < 3 \cdot 10^{-3}$,
the day-night effect is negligible.

%%%%%%%%%%%%%%%%%%%%%%%%%%%%%%%%%%%%%%%%%%%%%%%%%%%%%%%%%%%%%%%%%%%%%%%%%

\vskip 0.3truecm
\leftline {\bf 7. Distortion of the boron neutrino spectrum and signals
in}
{\bf SuperKamiokande and SNO}
\vskip 0.2truecm

An interplay of vacuum oscillations and resonance conversion can lead to
peculiar distortion of the boron neutrino energy spectrum.
In particular, for $\Delta m^2_{12} > 10^{-11}$ eV$^2$
one expects  an additional oscillatory modulation of the spectrum due to
vacuum oscillations (fig. 1).

Let us consider  a manifestation of such an oscillatory
distortion in the energy
spectrum of the recoil electrons measured by  the Super-Kamiokande
as well as (in future)  SNO    experiments.

In presence of the neutrino conversion described by the
survival probability $P(E_{\nu})$
the  number of recoil electrons, $N(E_{vis})$,
with a given visible
energy,  $E_{vis}$,  equals:
\begin{equation}
\begin{array}{c}
{\displaystyle N(E_{vis})~=~\int dE_e \cdot f(E_{vis},~E_e) \cdot
\int_{E_e - {m_e \over 2}} dE_{\nu} \cdot \Phi (E_{\nu})\cdot
~~~~~~~~~~~~~~~~~~~~~~~~~~~~~~~~~~~~~~~~~~~~}$$
\\$$~~~~~~~~~~~~~~~~~~~~~~~~
{\displaystyle
\left[
P(E_{\nu}) {d \sigma_{\nu_e} \over
{dE_e }}(E_e,~E_{\nu})~ +
(1 - P(E_{\nu}))
{d \sigma_{\nu_{\mu}} \over {dE_e }}(E_e, ~E_{\nu})
\right] ~},
\label{spectrum}
\end{array}
\end{equation}
where $E_e$ is the total energy of the recoil electron,
$\Phi (E_{\nu})$ is  the original boron neutrino flux,
$f(E_{vis}, E_e)$ is the energy resolution function
which can be parameterized as
\begin{equation}
{\displaystyle
f(E_{vis}, E_e)~=~
{1\over {\sqrt{2\pi} E_e \sigma(E_e)}}
\cdot
exp \left[
-\left( {E_{vis} - E_e \over {\sqrt{2}E_e\sigma (E_e)}}
\right)^2
\right ]
}.
\label{resolutionf}
\end{equation}
We use  the value of
$\sigma$ determined in the  Super-Kamiokande
calibration experiment \cite{SK1,SK2}. It
has an  approximate dependence on the neutrino energy
$\sigma \propto {1 \over \sqrt{E_e}}$.

Similar number of events
$N_0(E_{vis}) =
N(E_{vis}, P = 1)$ has been found  in the absence of conversion.
Using $N$ and $N_0$ we  calculate  ratios, $R^i_e$, of
events with and without oscillations in the $\Delta E = 0.5$ MeV energy
bins  corresponding to the Super-Kamiokande presentation:
\begin{equation}
{\displaystyle
R_e^i =
\frac{
\int_{E_i}^{E_i + \Delta E} dE_{vis}' N(E_{vis}')
}
{
\int_{E_i}^{E_i + \Delta E} dE_{vis}' N_0(E_{vis}')}
}~.
\label{ratio}
\end{equation}
In figs. 3, 4 we show (by histograms) the ratios $R_e^i$ expected for
different values of the
neutrino parameters as well as the ratios
measured by the Super-Kamiokande experiment during 306 days \cite{SK2}.
For comparison we present also the distortions expected from
the pure MSW-solution (fig. 3a) and pure vacuum oscillation
solution (fig. 3b).

As follows from fig. 4, the integrations over the neutrino energy
and the electron
energy convoluted  with the resolution function lead to  strong averaging
of the oscillatory behavior. Indeed, for present water Cherenkov
detectors, like the Super-Kamiokande, the width of the
resolution function (on semi-height), $2 \sigma$,  is  $\sim 4$ MeV
at 10 MeV  which is  comparable with the width of the largest dip
of the oscillatory curve (in the energy scale).
To illustrate the effect of smoothing  we show
the ratios for the ideal resolution
$
f(E_{vis}, E_e) = \delta (E_{vis} - E_e)~$.
The most profound effect follows from the first (the widest)
dip of the oscillatory curve.

Let us consider manifestations of the oscillatory
behavior in details. For this  we  fix parameters of the
$\nu_e - \nu_{\tau}$ system and find the modification of the
dependence of  $R_e$ on $E_{vis}$
for different values  $\Delta m^2_{12}$.
We will fix $\sin^{2} 2\theta_{e \mu} = 0.5$.
An increase of the angle $\theta_{e\mu}$ leads to enhancement of distortion.

With increase of
$\Delta m_{12}^2$ the oscillatory curve which corresponds
to the vacuum oscillation
probability shifts to  high energies (see fig. 1). In such a way
different parts of the oscillatory curve turn out to be
in the observable range  of the boron neutrino ($\nu_B$-)  spectrum:
$E \sim (5 - 15)$ MeV.

We find the following distortion picture for  different
intervals of $\Delta m^2$:

\begin{itemize}

\item
$\Delta m_{12}^2 < 10^{-11}$ eV$^2$:
The oscillatory curve is below the observable range
$(5 - 15)$ MeV.
Therefore the   distortion of $\nu_B$-spectrum
is close to that  produced by the  resonance conversion
alone.

\item
$\Delta m_{12}^2 < 6 \cdot 10^{-11}$ eV$^2$:
the $\nu_B$-spectrum
is on the rising (with energy) part of the first (high energy) dip
of the oscillatory curve.  In this case the vacuum oscillations
 enhance the distortion produced by  the resonance conversion.
The slope of the $R_e (E_{vis})$  is bigger
than in the pure MSW-case (fig. 4 a).

\item
$\Delta m_{12}^2 =  (6 - 10)  \cdot 10^{-11}$ eV$^2$:
the minimum of the
first dip  of the oscillatory curve is in the range
$(5 - 15)$ MeV.
This leads to the
dip in the  $R_e (E_{vis})$  dependence (fig. 4 b, c).

\item
$\Delta m_{12}^2 =  (1.0 - 1.2)  \cdot 10^{-10}$ eV$^2$:
the $\nu_B$-spectrum is in the decreasing part of the first dip
which  manifests itself  as the
negative slope in the distortion of
$R_e (E_{vis})$ (fig. 4d).

\item
$\Delta m_{12}^2 =  (1.2 - 2)  \cdot 10^{-10}$ eV$^2$:
the first (high energy) bump of the oscillatory curve is in the
range $(5 - 15)$ MeV which produces the bump in the
recoil electron spectrum (fig. 4e, 4f).

\item
$\Delta m_{12}^2 =  (2 - 5)  \cdot 10^{-10}$ eV$^2$:
the second dip of the oscillatory curve
is in the range  $(5 - 15)$ MeV. Again  with
increase of   $\Delta m_{12}^2$   first an enhancement of the slope
occurs (fig. 4g),  then a dip appears, and
finally the slope becomes negative
(fig. 4h).

\item
$\Delta m_{12}^2 > 5  \cdot 10^{-10}$ eV$^2$:
the second dip shifts to $E > 15$ MeV.
There are several narrow dips in the observable
part of spectrum. Some non-averaged effect still exists in
the high energy part.

\item
$\Delta m_{12}^2 > 10^{-9}$ eV$^2$:
The periods of the oscillatory  curve in the observable interval are
substantially smaller
than $2\sigma$.
Strong averaging of the oscillation effect takes place. The
ratio $R_e$ approaches the asymptotic dependence which
can be obtained from (\ref{spectrum}) by substitution
$P_V \rightarrow \bar P_V = 1 - 0.5 \sin^2 2\theta_{e \mu}$:
\begin{equation}
R_e \approx R_e^R \cdot \bar P_V +
(1 - \bar P_V) \frac{N_{\mu}}{N_0} =
(1 - 0.5 \sin^2 2\theta) \cdot R_e^R +
0.5 \sin^2 2\theta \frac{N_{\mu}}{N_0}~,
\label{asymptot}
\end{equation}
where $R_e^{R}$ is the ratio for the case
of pure resonance conversion, $N_{\mu}$ is the effect of
muon (tau)-neutrino:
\begin{equation}
{\displaystyle
N_{\mu} \equiv  \int dE_e \int dE_{\nu} f \cdot  \Phi \cdot
        {d \sigma_{\nu_{\mu}} \over {dE_e }}
}~.
\label{numu}
\end{equation}
Note that this contribution does not depend on the
probabilities of transitions.
$N_0$ is the number of events in absence of oscillations.
The last term in (\ref{asymptot}) is relatively small:
$\sim 0.04$.

According to (\ref{asymptot}) for  large
$\Delta m_{12}^2$, vacuum oscillations result in flattening
of the distortion stipulated  by  the resonance conversion alone
(decrease of the slope).

\end{itemize}

In certain intervals of $\Delta m_{12}^2$
an additional effect of vacuum oscillations leads to
distortion which contradicts  observations.
Therefore, in these intervals  large mixing angles (typically
$\sin^2 2\theta_{e \mu} > 0.5 $) can be already excluded.

The smoothing effect is weaker in the SNO experiment.
The integration over the neutrino energy  gives
weaker averaging and energy resolution is expected to be slightly
better. In our calculation we use the cross-sections of reaction $\nu
d ~\rightarrow~epp$ from \cite{dcross} and  the energy resolution
function in the form (28) with $\sigma(E_e)~=~14\%/
\sqrt{10 {\rm MeV}/ E_e }$.
One needs experiments with at least two times better energy resolution
(like HELLAZ \cite{HEZ}) to measure the  oscillatory
modulation of the energy spectrum.

Deviations of the spectrum distortion from the
simple form predicted by the pure  MSW solution or vacuum oscillation solution
will  indicate  the effect of the third neutrino considered above.

%%%%%%%%%%%%%%%%%%%%%%%%%%%%%%%%%%%%%%%%%%%%%%%%%%%%%%%%%%%%%%%%%%%%%%%%%

\vskip 0.3truecm
\leftline {\bf 8. Seasonal variations of signals}
\vskip 0.2truecm

The vacuum oscillation probability
depends explicitly on the distance
between the Sun and the Earth $L$:
$P_V =  P_V(L)$.
Therefore seasonal variations due to pure geometrical effect,
$1/L^2$, related to the eccentricity of the Earth
orbit will be modified \cite{season}. Depending on values of the
neutrino parameters, the geometrical effect can be
enhanced or suppressed by the $L$-dependence in the probability.
Moreover, for experiments which are sensitive to
the high energy part of the boron neutrino spectrum
the modification is strongly
correlated to the distortion of the recoil
electron energy spectrum \cite{seasonms}.

Let us consider   the seasonal variations in the case of
hybrid solution. Using 
the probability (\ref{Pf}) we can rewrite
the expression for the number of events in the
Super-Kamiokande experiment (\ref{spectrum})
in the following way:
\begin{equation}
{\displaystyle N = \int dE_e \int dE_{\nu} f \cdot  \Phi \cdot
P_R \cdot P_V \cdot
\left[  {d \sigma_{\nu_e} \over {dE_e }} -
        {d \sigma_{\nu_{\mu}} \over {dE_e }}
\right]  + N_{\mu}
}~,
\label{spectrum1}
\end{equation}
where $N_{\mu}$ is defined in (\ref{numu}).
Note that $N_{\mu}$ changes with  time only due to
the  geometrical dependence of  the flux.
If the resonance conversion probability
varies weakly in the
observable energy interval, we can substitute it by certain
average value $\bar P_R$. Then the   expression for
the number
of event becomes
\begin{equation}
N  \approx \bar P_R \cdot N_{vac} + (1 - \bar P_R)\cdot N_{\mu}~,
\label{rate1}
\end{equation}
where $N_{vac}$ is the number of events expected in the
case of pure vacuum oscillations.  Notice that for the configuration shown
in  the fig.1 the probability $P_R$ indeed changes in a small interval
from 0.7 to 0.9. According to (\ref{rate1}) the vacuum oscillation
effect is attenuated by the factor $\bar P_R$. This means that to get the
same overall suppression of flux one needs smaller value of
$\sin^2 2\theta_{e\mu}$,  and therefore the
seasonal variations become weaker. Moreover, there is
an additional  damping term in (\ref{rate1}).

Let us introduce the seasonal asymmetry $A$ as
\begin{equation}
A = 2 \frac{N^W - N^S}{N^W + N^S}~,
\label{asymm}
\end{equation}
where $N^W = \int_W dt N(t)$ and
$N^S = \int_S dt N(t)$
are  the integral numbers of events during the winter and
summer time  correspondingly.
Using (\ref{rate1}) we get
\begin{equation}
A =  A_V \cdot  \left[1 -
(1 - \bar P_R )\cdot \frac {N_{\mu}}{\bar N} \right]~.
\label{asymm1}
\end{equation}
Here $A_V$ is the asymmetry in the case of pure vacuum oscillations
and $\bar N$ is the number of events averaged over the year.
The last term in (\ref{asymm1}) is typically smaller than 10\%.
As we mentioned above,  in  presence of the resonance conversion
the value  $\sin^2 2 \theta_{e\mu}$
which leads to a good fit of the data is smaller than in the
pure vacuum oscillation case:
$\sin^2 2 \theta_{e\mu} < 0.8$. Then from (\ref{asymm1}) we find that
seasonal variations related to the probability are  attenuated by
some  factor
$ < 0.7$ in comparison with variations in the  pure vacuum oscillations
solution.

%%%%%%%%%%%%%%%%%%%%%%%%%%%%%%%%%%%%%%%%%%%%%%%%%%%%%%%%%%%%%%%%%%%%%%%%

\vskip 0.3truecm
\leftline{\bf 9. Discussion and conclusions}
\vskip 0.2truecm

Let us first summarize our main results.

1. We have considered the Grand Unification scenario
which is based on the see-saw mechanism
with the Majorana mass of the RH neutrino
$M_{3} \sim \Lambda_{GU}$ and on the quark-lepton symmetry for  fermions
from the third generation.  The $GU$-scenario leads to the mass of the
third neutrino, $m_3 \sim (2 - 4) \cdot 10^{-3}$ eV,  in the range
of masses implied by the  resonance conversion
of the solar neutrinos.

2. We have found that in the $GU$-scenario the MSW - solution
of the $\nu_{\odot}$-problem implies  the mass
$M_3 \sim (0.2 - 0.7)\Lambda_{GU}$
(depending on the value of $\tan\beta $) in the absence of mixing of the RH
neutrinos and $M_3$ can coincide with  $\Lambda_{GU}$ if small
($\theta_M \sim \sqrt{M_1/M_3}$)  mixing in the Majorana
mass matrix of the RH neutrinos is taken into
account.

The proximity of the mass $M_3$ to
$\Lambda_{GU}$ can be considered as an
indication of the Grand Unification.

3. Assuming a linear  hierarchy
of the RH neutrino masses or/and the presence of
contributions from non-renormalizable operators related to Planck scale
physics  we get   $m_2 \sim (0.3 - 3)~10^{-5}$ eV$^2$ and
$\Delta m^2_{12}\sim
(10^{-11} -  10^{-9}$) eV$^2$. Furthermore,
the mixing between the first and the second generations
can be rather large:
$~\sin^2 2\theta_{e\mu} \stackrel{_>}{_\sim} 0.2$.
In this case the \nue $\rightarrow$ \numu  vacuum oscillations
on the way from the Sun to the Earth  give rise to
observable effects.

4. The $GU$-scenario leads to  non-trivial
interplay of the
\nue $\rightarrow$  \nutau  resonance conversion and
\nue $\rightarrow$ \numu vacuum oscillations of the solar neutrinos.
An additional vacuum oscillation effect enlarges possible range of
neutrino mixing,  so that  $\sin^2 2\theta_{e\tau}$
 can be as small as $2 \cdot 10^{-4}$ which is
of the order of corresponding quark mixing.

5. The interplay of the resonance conversion and vacuum oscillations
leads to new type of distortion of the boron neutrino spectrum, namely,
to oscillatory modulation of the energy dependence produced by
the resonance conversion alone.

The integrations over the neutrino energy and
the electron energy convoluted with the  energy resolution function
in the  Super-Kamiokande experiment (and similarly SNO) result in strong
smoothing of the oscillatory behavior  of the recoil electron energy
spectrum. Still  depending on  $\Delta m^2_{12}$
one can observe an enhancement of the positive slope, or  appearance of
negative slope as well as dip or bump in the
dependence of $R_e$ on $E_{vis}$. Two times better energy resolution is
needed to observe real oscillatory behaviour.

Certain ranges of parameters
are already excluded by existing  data.

6. Seasonal  variations of  signals are expected due to 
dependence of the vacuum oscillation probability on 
distance from the Sun to the Earth. Typically these
variations are expected to be weaker than in the pure vacuum oscillation
case.

The $GU$- scenario considered in this paper does not allow one to explain the
LSND result or to solve the atmospheric neutrino problem
in terms of neutrino oscillations.
It predicts ``null" results in the CHORUS and NOMAD experiments
as well as in future short and long baseline oscillation experiments.
Null result is also expected in the neutrinoless double beta decay
searches etc.
Therefore a positive result in at least one of  these experiments will
exclude
the simplest version
of the $GU$-scenario.
Also the scenario  does not supply any appreciable contribution to
the hot dark matter (HDM) in the Universe.

 It is possible to modify the $GU$-scenario discussed here so 
that the second
neutrino mass is $m_2 \sim (0.1 - 1)$ eV.
For this the mass of the corresponding  RH neutrino should be
$M_{02} \sim 10^{10}$ GeV in the absence of mixing
(or for small mixing).
%(The matrix has the structure $M_R \approx diag(0, 0,  \Lambda_{GU})$).
Such a modification can accommodate the LSND result and give a
sizable  contribution to HDM.
An explanation of the atmospheric neutrino anomaly
requires also large mixing between $\nu_{\mu}$
and $\nu_{\tau}$. This is more difficult to achieve, since
strong mass hierarchy accompanied by large mixing angle
implies  further tuning of parameters.

Additional freedom in explaining of the data arises
if  new (sterile) neutrino states are introduced.\\

\vglue 0.4cm
\leftline{\bf Acknowledgments.} The authors are grateful
to F. Vissani for discussions concerning renormalization group effects.
The work of K.S.B. is supported by a grant from DOE, Q.Y.L.
is supported in part
by the EEC grant ERBFMRXCT960090.  K.S.B. wishes to acknowledge the warm
hospitality he received from the Theory Group at
LBL where part of this work was done.  He also wishes to thank C. Kolda and
P. Krastev for discussions.  Q.Y.L. would like to acknowledge financial
help from {\it 2nd International Symposium on Symmetries in Subatomic Physics}
at the University of Washington where a draft of this paper was presented.

\vglue 0.4cm
\leftline{\bf Note added.}
While this work was revised the paper \cite{PARIDA} appeared in which
renormalization effects for neutrinos in a similar context have been
studied. Our results are in agreement with those in \cite{PARIDA}.

\begin{table}
\caption[] { The mass of third neutrino $m_3$ as the function of
$\tan\beta$
for \MR$=10^{16}$ GeV and three possible values of physical top quark mass:
$m_t$ = 175, 170, 180 GeV.

\protect\label{taumass}}
\begin{tabular}{l c c c}
tan$\beta$ &  & \nutau -mass ($10^{-3}$ eV) &   \\ \hline
 &  $m_t=170$GeV &  $m_t=175$GeV  &  $m_t=180$GeV  \\ \hline
1.7   & 1.47 &  2.85 & --- \\
1.9   & 1.16 &  1.64 & 2..71 \\
3  & 0.83 &  0.99 & 1.17 \\
5  & 0.76 & 0.88 & 1.00 \\
10  & 0.73 & 0.84 & 0.95 \\
%20  & 0.73 & 0.84 &  0.95 \\
30  & 0.74 & 0.85 & 0.96 \\
%40  & 0.76 & 0.86 & 1.00 \\
50  & 0.76 & 0.92 & 1.05 \\
60  & 0.83 & 0.99 & 1.17 \\
%63  & 0.84 & 1.01 & 1.21
\end{tabular}
\end{table}

\newpage

\begin{figure}[H]
\mbox{\epsfig{figure=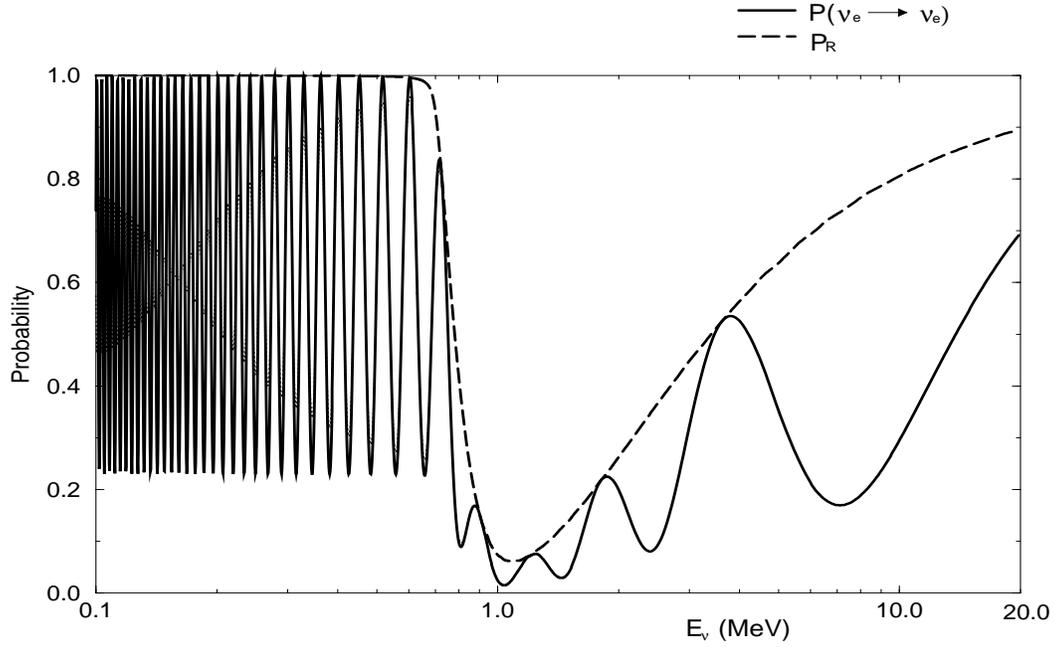,width=10cm,height=14cm,angle=-90}}
\vglue0.5cm
\caption[]{The averaged (over production area) survival probability
$P(\nu_{e} \rightarrow \nu_{e})~$ (solid line) as a function
of the neutrino energy $E$ for the following values of
parameters: $\sin^22\theta_{e\mu}=0.77$, $\Delta
m^2_{21}=6.0 \cdot 10^{-11}\rm{eV}^2$, $\sin^22\theta_{e\tau}=8.0\cdot
10^{-4}$, $\Delta m^2_{31}=1.1 \cdot 10^{-5} \rm{eV}^2$. The
dependence of the averaged MSW probability $P_R(\nu_{e}
\rightarrow \nu_{e})~$ alone on $E$ is also shown (dashed line). }
\end{figure}

\begin{figure}[H]
\mbox{\epsfig{figure=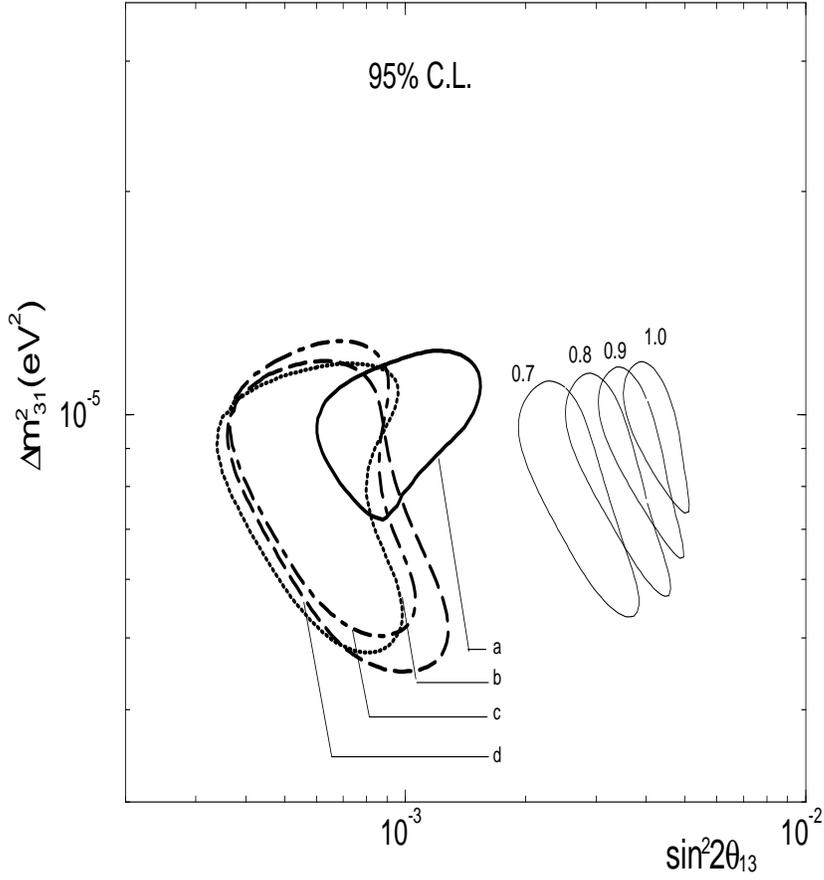,width=12cm,height=14cm}}
\vglue0.5cm
\caption[]{The 95\% C.L. regions of solution of the $\nu_{\odot}$-problem
in  $\Delta m^2_{31}$ and $\sin^22\theta_{e\tau}$ plot
(306 days Super-Kamiokande data
are taken into account) for different vacuum
oscillation parameters
$\sin^22\theta_{e\mu}$, $\Delta m^2_{21}$, and the original boron neutrino
flux (in units of the standard solar model)
$f_B$ $\equiv {\displaystyle {\Phi(^8B) / {\Phi_{SSM}(^8B)}}}$.
The thin solid lines correspond to $\sin^22\theta_{e\mu}=0.48$,
$\Delta m^2_{21}=1.4 \cdot 10^{-10}$ eV$^2$, and the values of
$f_B$ indicated at the curves.
 The thick lines correspond to (a) $\sin^22\theta_{e\mu}=0.66$,
$\Delta m^2_{21}=8.3 \cdot 10^{-11}$ eV$^2$ and $f_B$=1.0;
 (b) the same as (a) but $f_B$=0.9;
 (c) $\sin^22\theta_{e\mu}=0.62$, $\Delta m^2_{21}=9.5 \cdot 10^{-11}$
eV$^2$, $f_B=0.8$;
 (d) $\sin^22\theta_{e\mu}=0.50$,
$\Delta m^2_{21}=7.8 \cdot 10^{-11}$ eV$^2$, $f_B=0.7$.
 }
\end{figure}

\begin{figure}[H]
\hglue 1.5cm
\mbox{\epsfig{figure=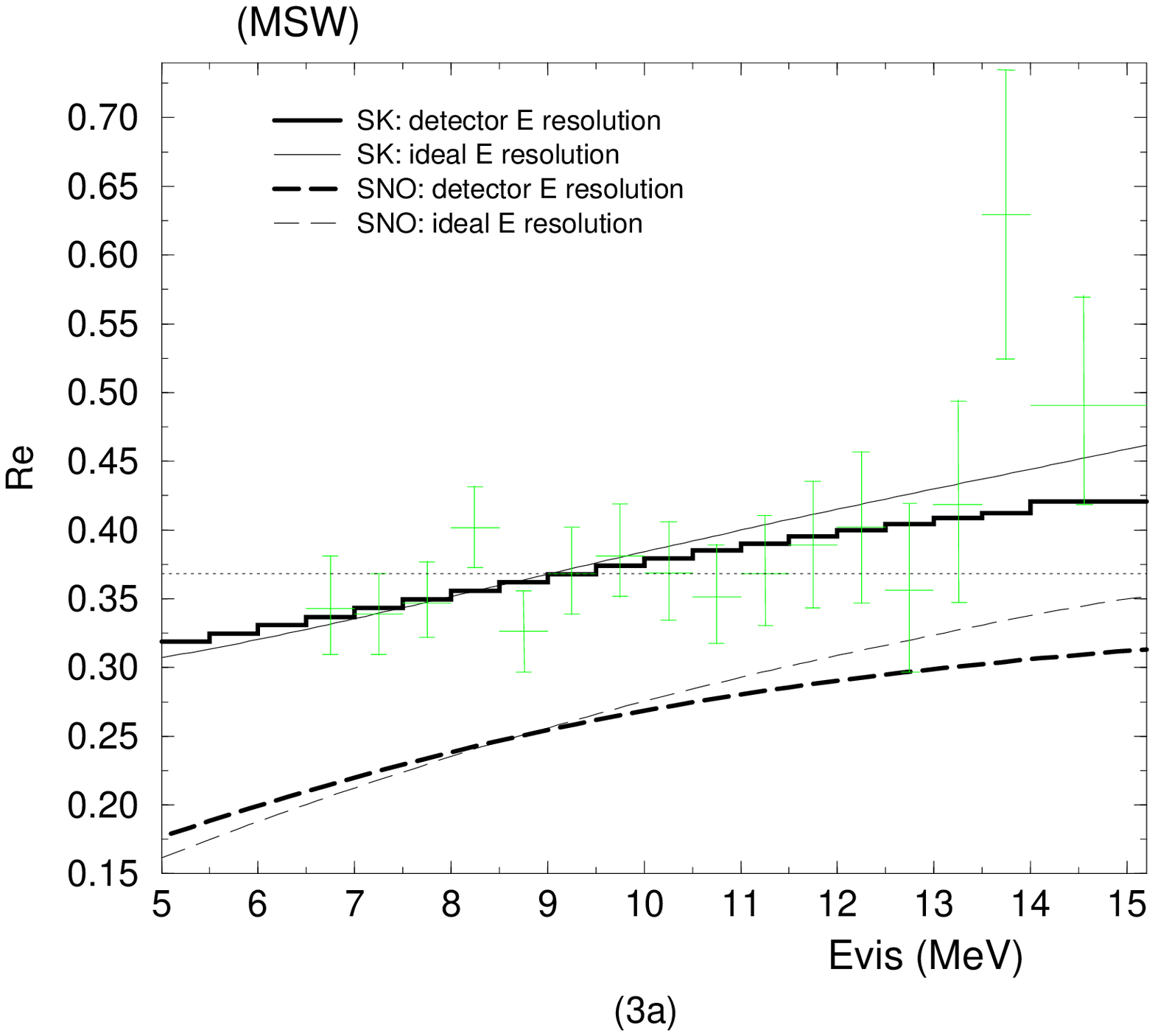,width=10cm,height=10cm}}
\vglue0.1cm
\hglue 1.5cm
\mbox{\epsfig{figure=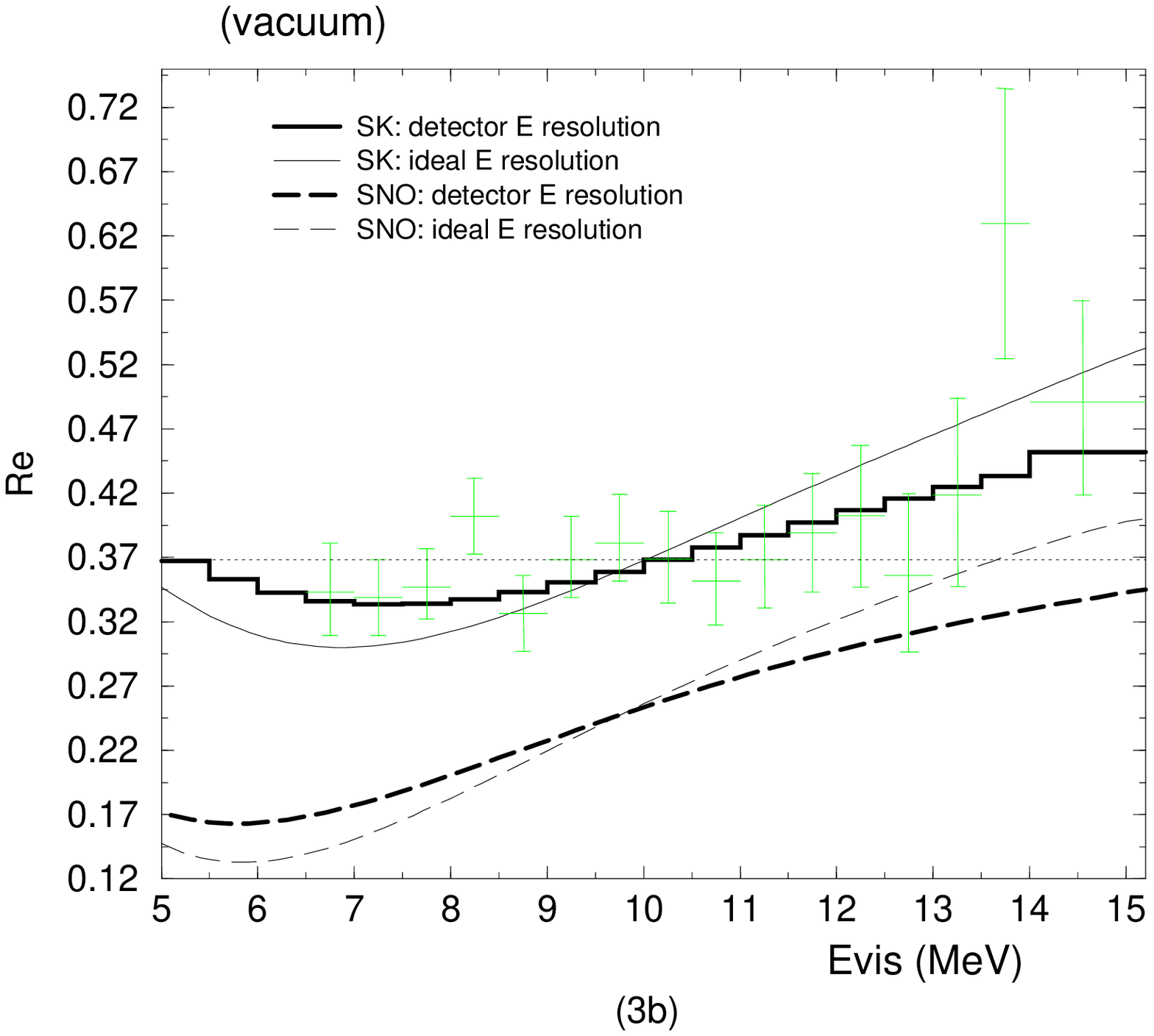,width=10cm,height=10cm}}
\vglue0.5cm
\caption[]{The expected spectrum deformations of the recoil electrons
in the Super-Kamiokande (solid lines) and
SNO (long dashed lines) experiments for (3a) two neutrino conversion with
$\sin^22 \theta=8.8 \cdot 10^{-3}$,
$\Delta m^2=5.0 \cdot 10^{-6}$ eV$^2$, (3b) vacuum
oscillation with $\sin^22\theta=0.82$,
$\Delta m^2=6.4 \cdot 10^{-11}$ eV$^2$.
The curves are normalized so that $R_e$(10MeV)=0.368 for the
Super-Kamiokande and $R_e$(10MeV)=0.256 for the SNO experiment. Thin lines
correspond to ideal
energy resolution; thick lines correspond to real energy resolution.
The Super-Kamiokande 306 days data are shown (statistical errors only). }
\end{figure}

\begin{figure}[H]
\hglue 1.5cm
\mbox{\epsfig{figure=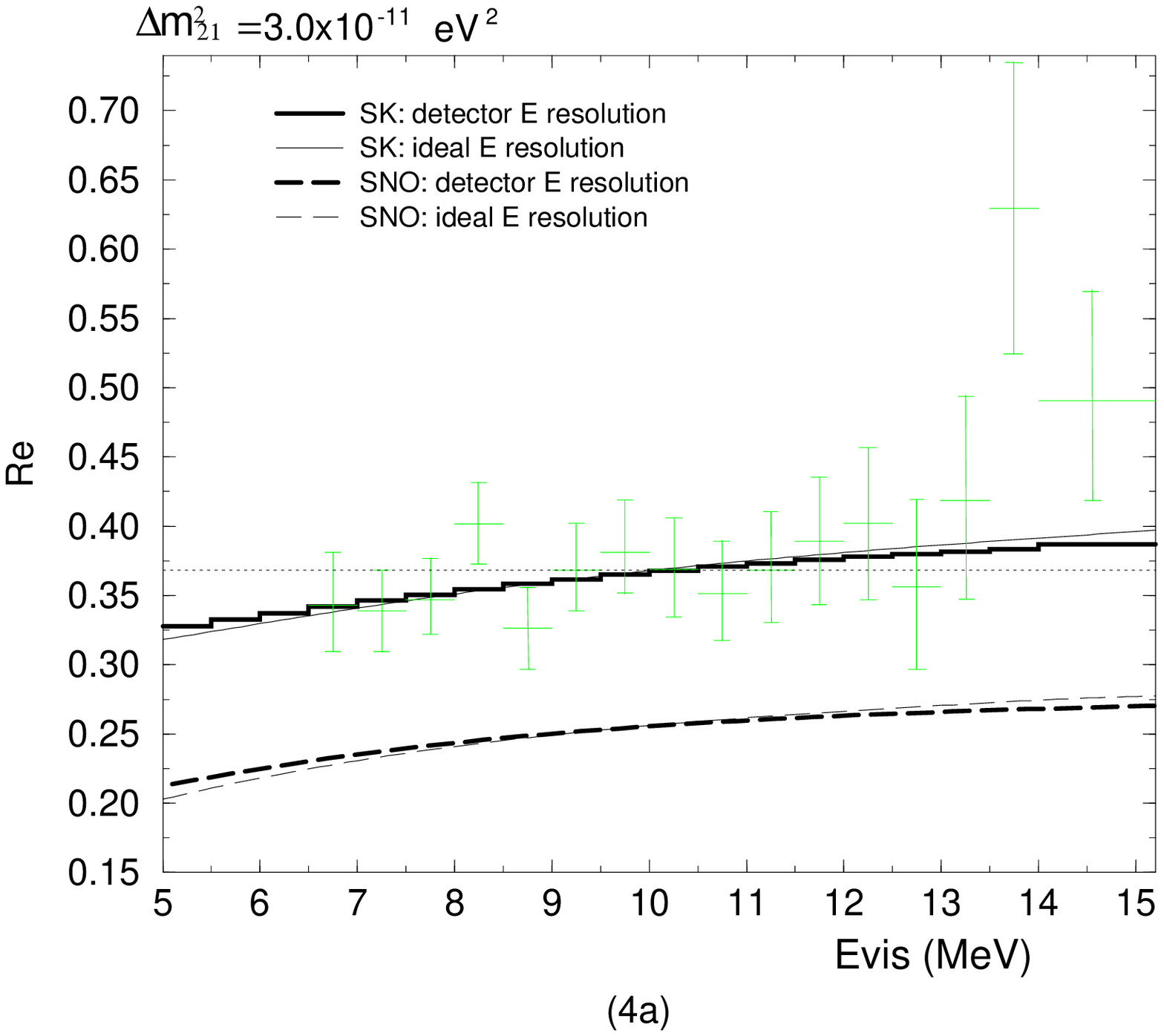,width=10cm,height=10cm}}
\vglue0.2cm
\hglue 1.5cm
\mbox{\epsfig{figure=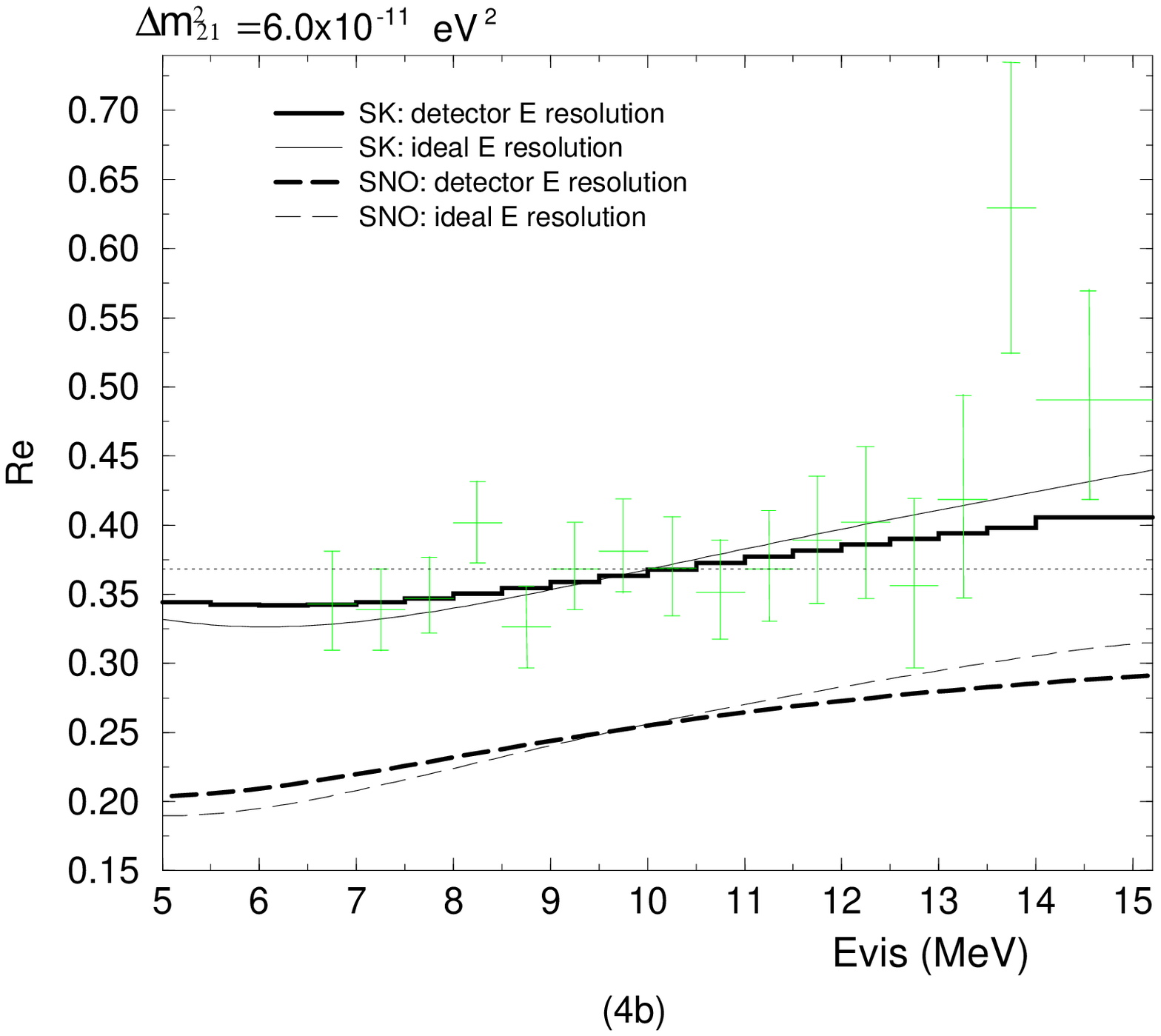,width=10cm,height=10cm}}
\vglue0.5cm

\newpage

\hglue 1.5cm
\mbox{\epsfig{figure=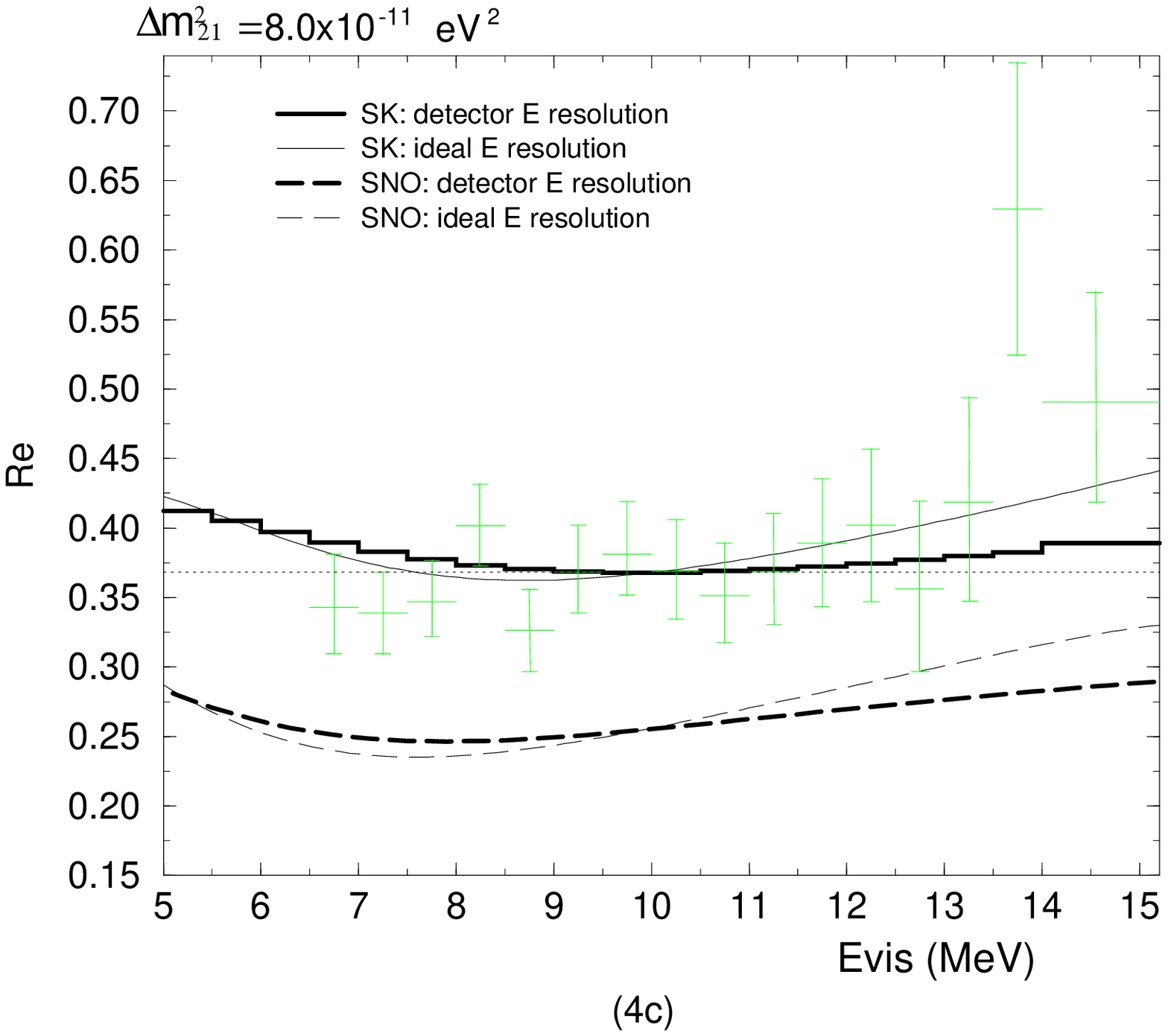,width=10cm,height=10cm}}
\vglue0.2cm
\hglue 1.5cm
\mbox{\epsfig{figure=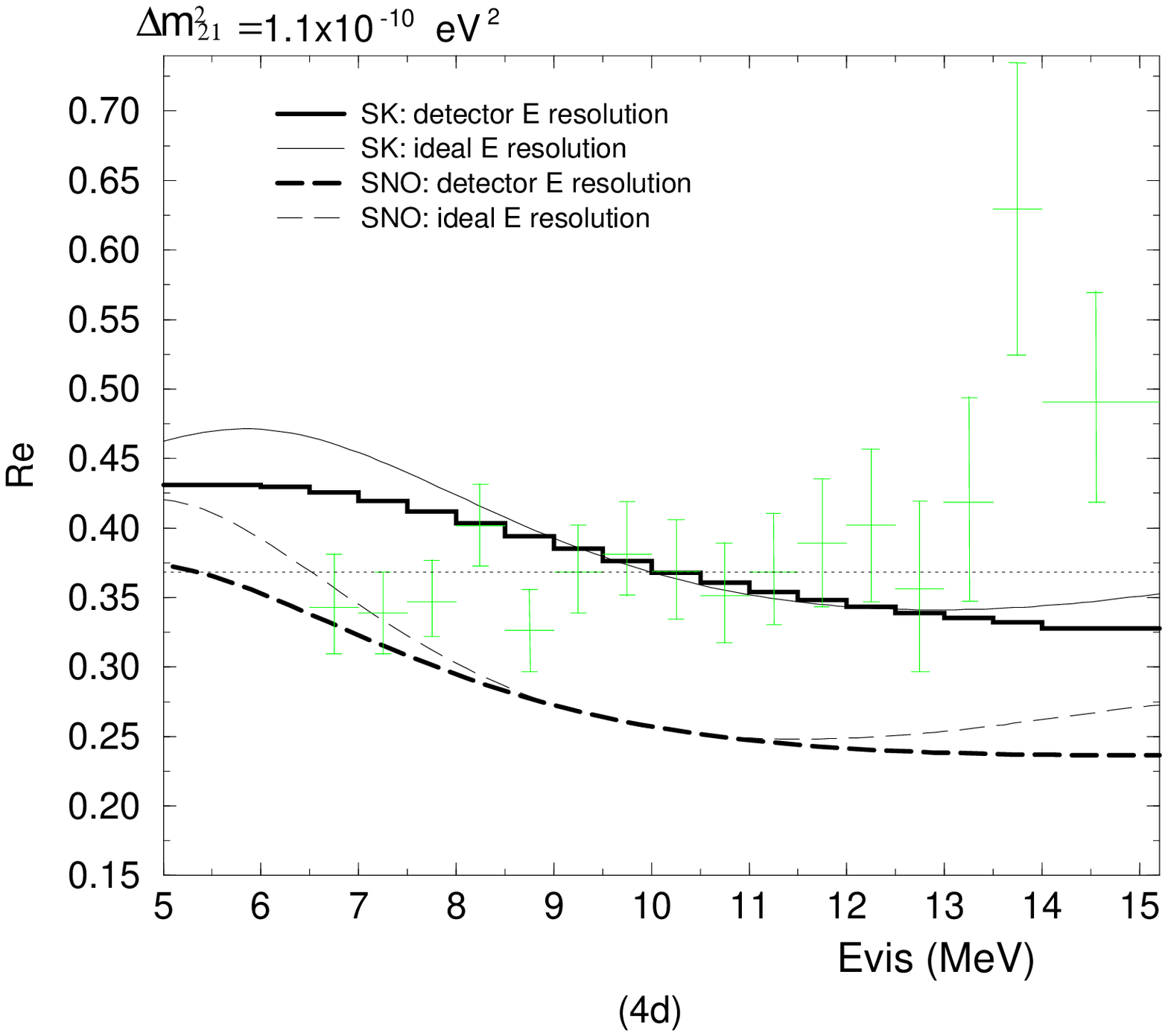,width=10cm,height=10cm}}
\vglue0.5cm

\newpage

\hglue 1.5cm
\mbox{\epsfig{figure=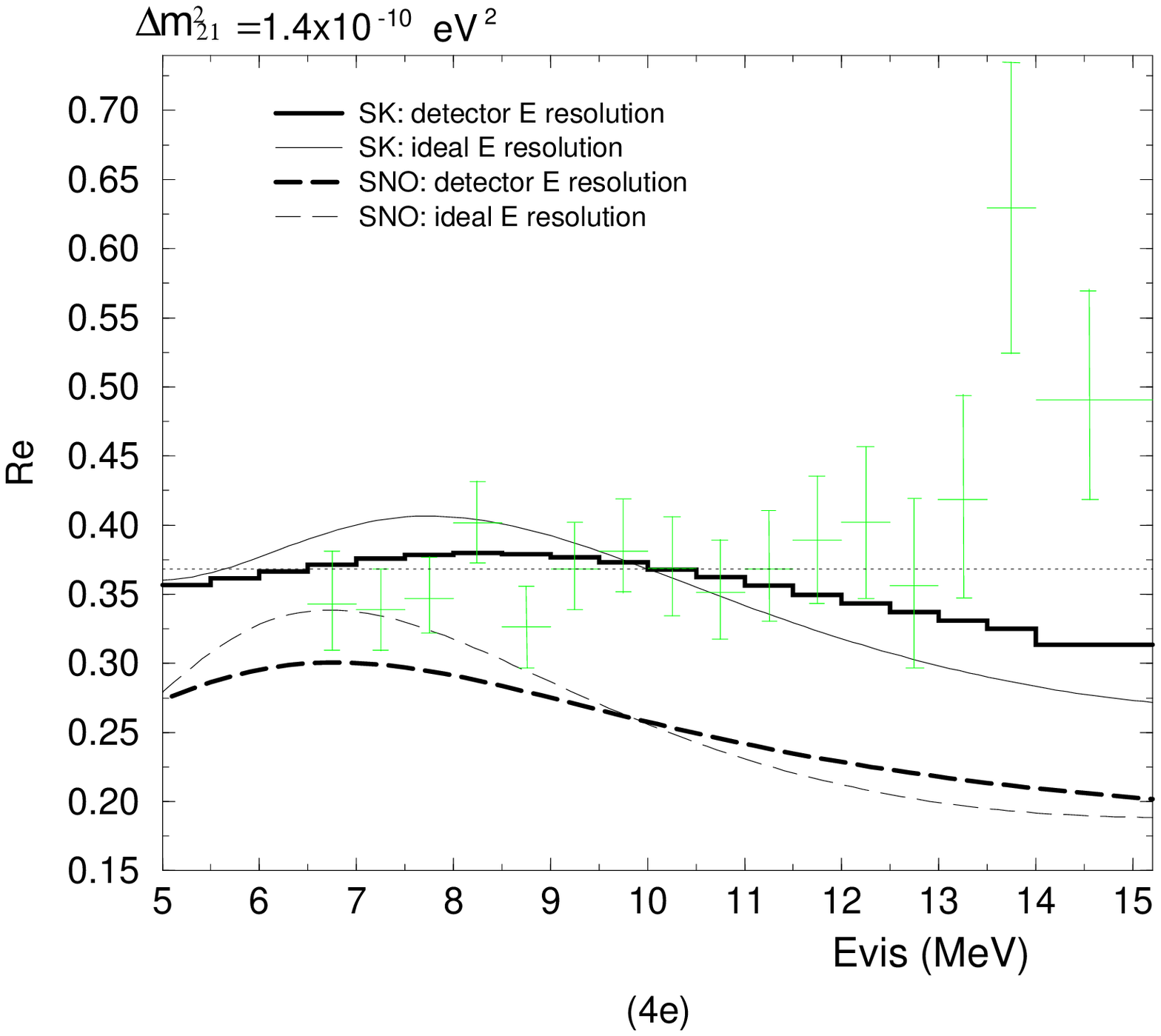,width=10cm,height=10cm}}
\vglue0.2cm
\hglue 1.5cm
\mbox{\epsfig{figure=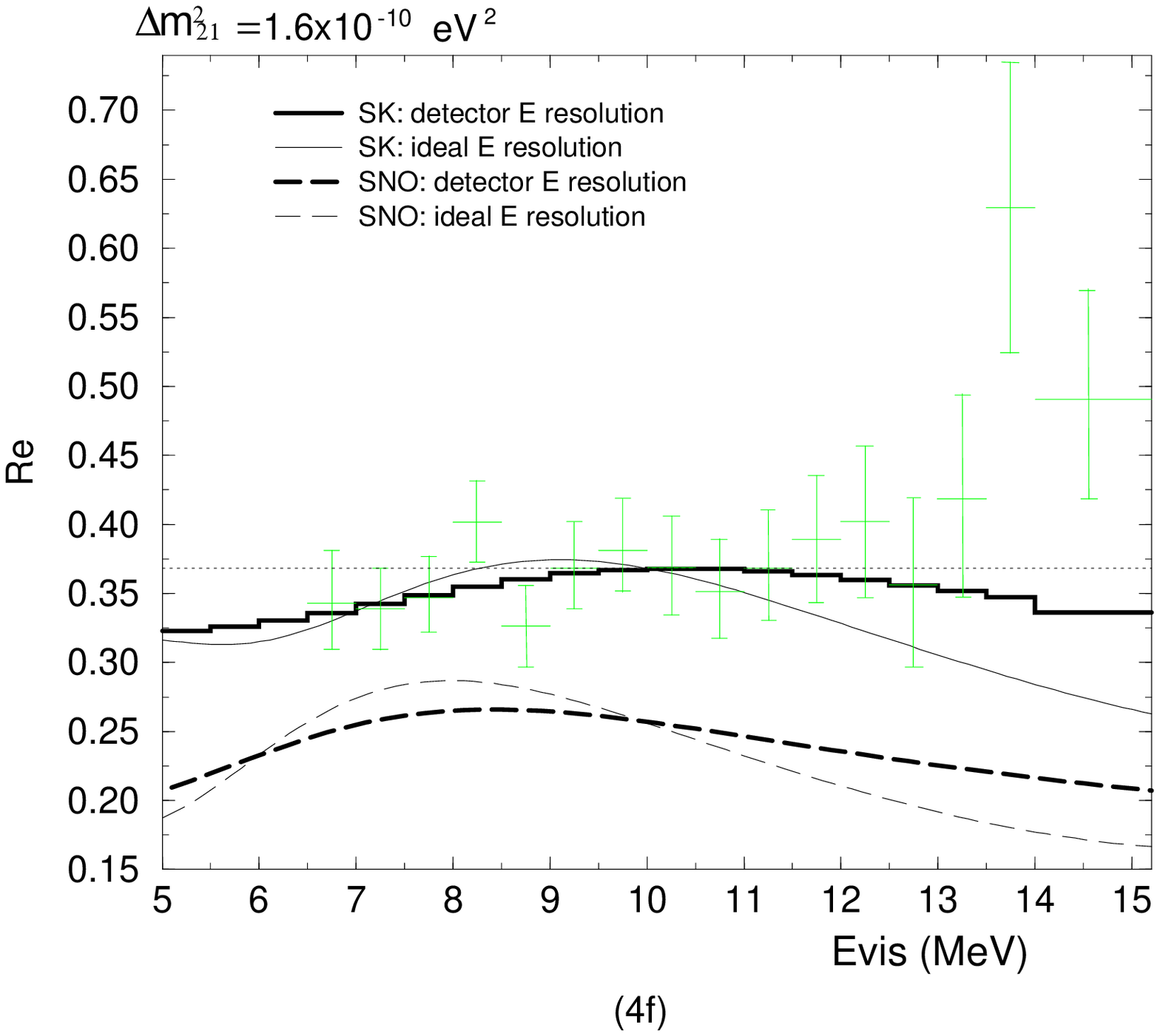,width=10cm,height=10cm}}
\vglue0.5cm

\newpage

\hglue 1.5cm
\mbox{\epsfig{figure=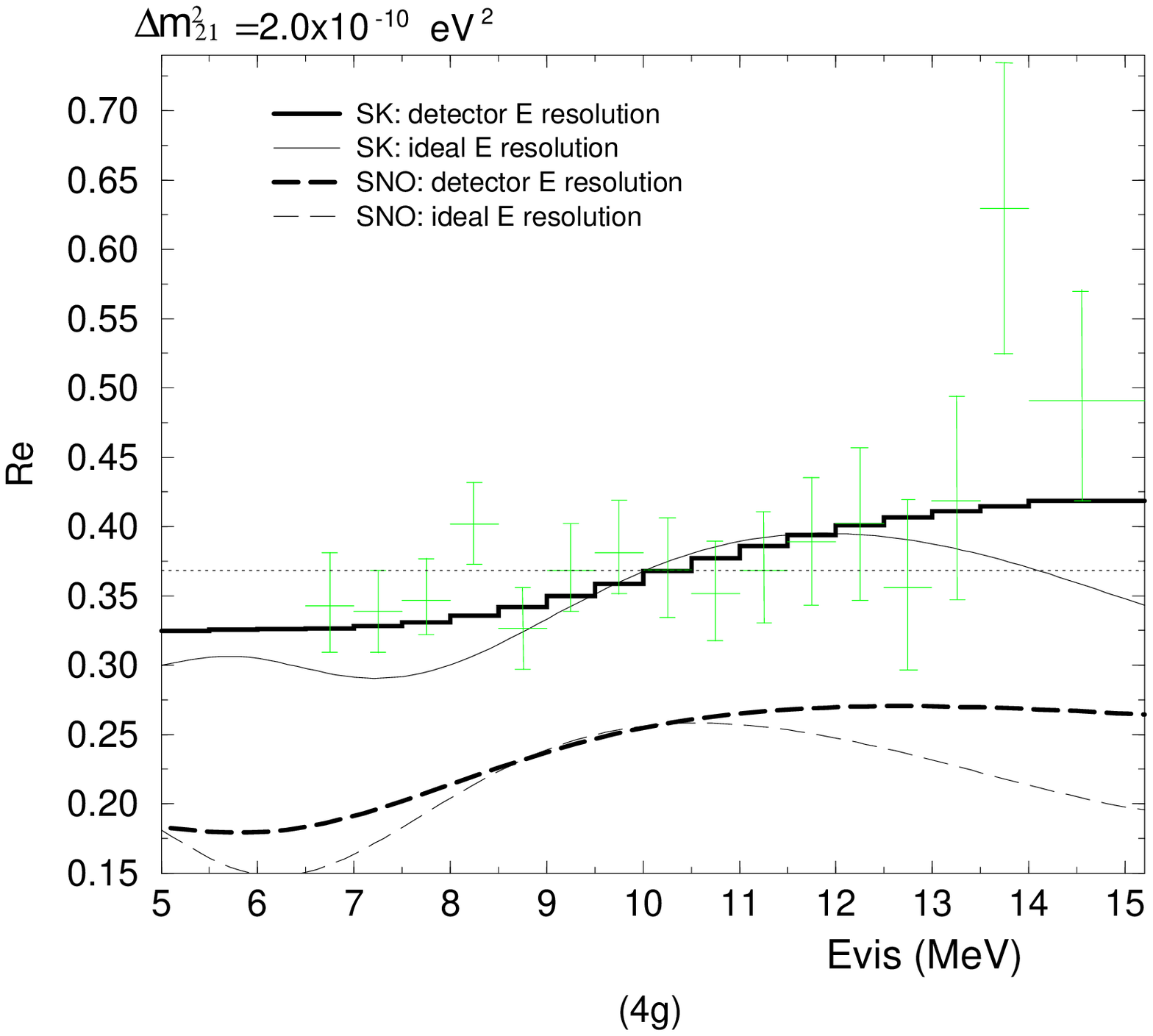,width=10cm,height=10cm}}
\vglue0.5cm
\hglue 1.5cm
\mbox{\epsfig{figure=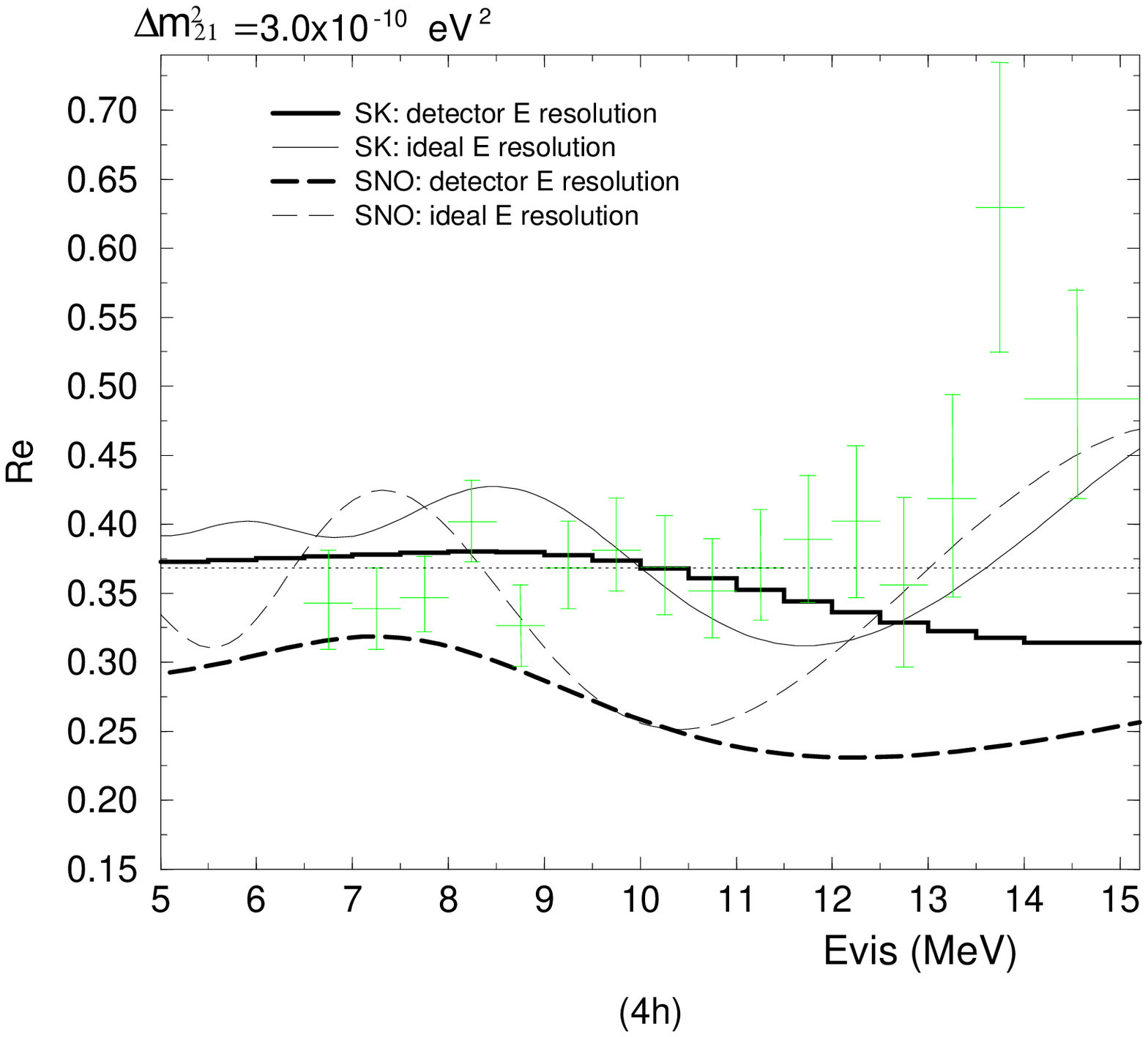,width=10cm,height=10cm}}
\vglue0.5cm

\newpage

\hglue 1.5cm
\mbox{\epsfig{figure=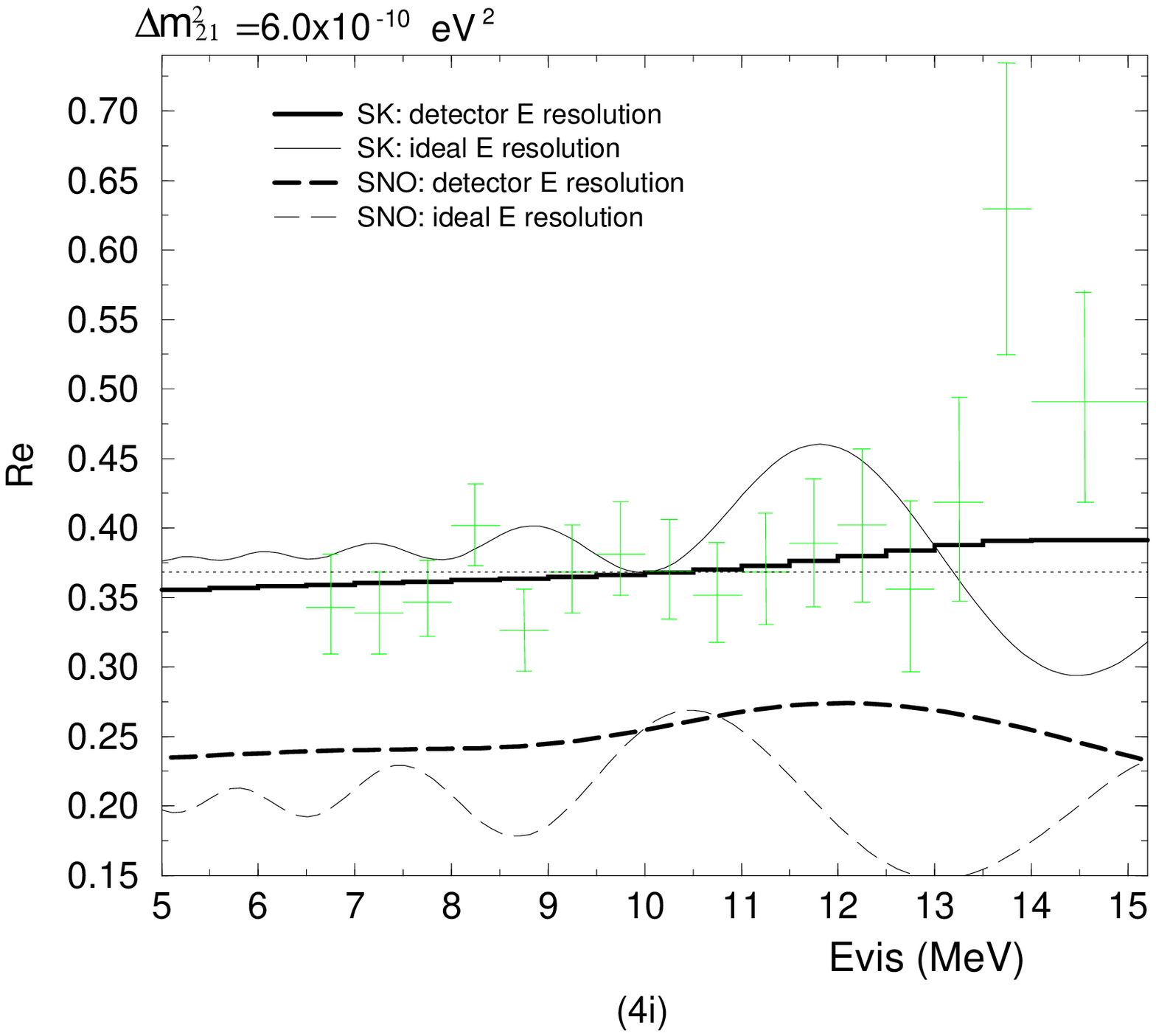,width=10cm,height=10cm}}
\vglue0.5cm
\hglue 1.5cm
\mbox{\epsfig{figure=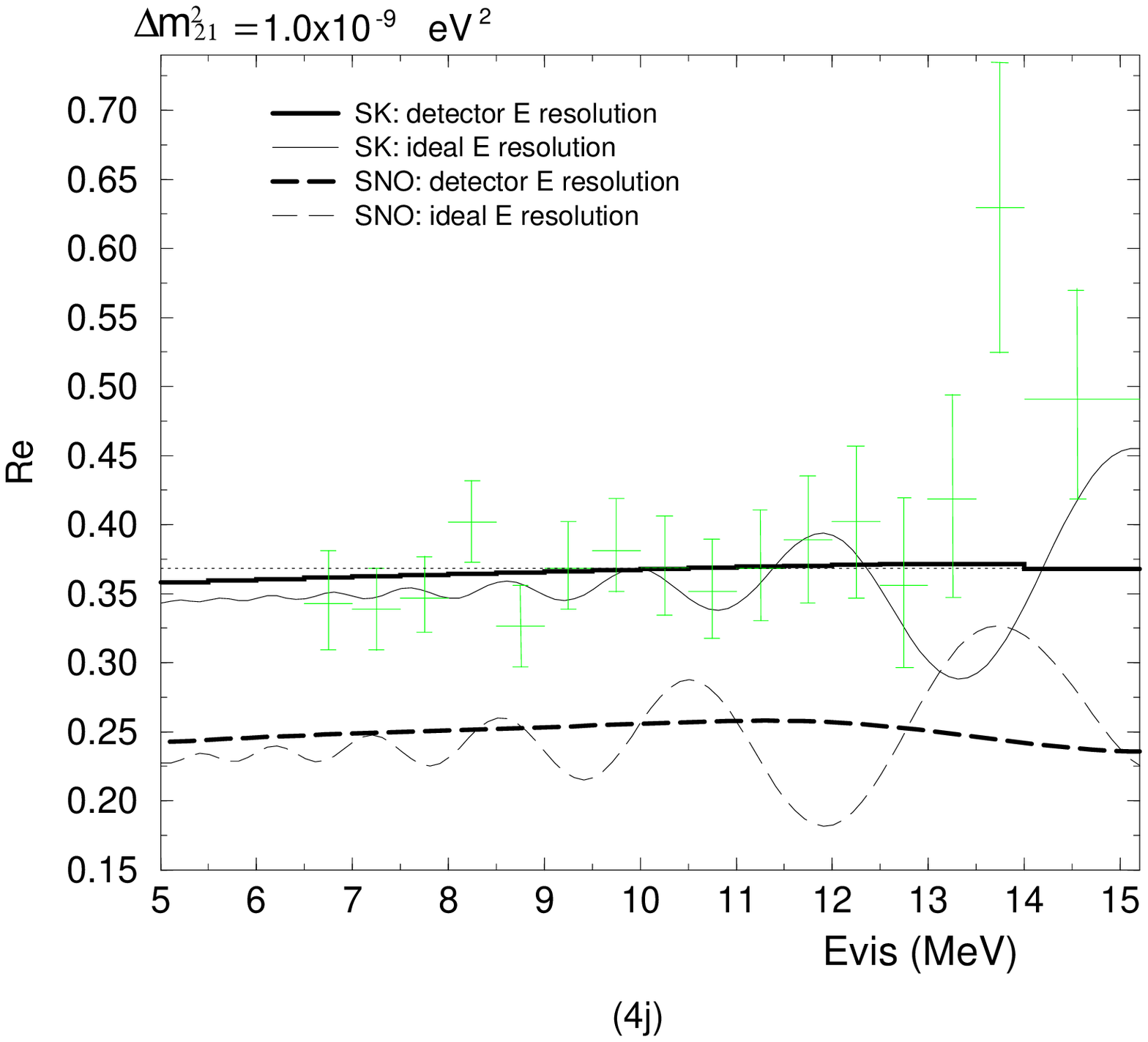,width=10cm,height=10cm}}
\vglue0.5cm
\caption[]{ The same as in figs. 3 but for the hybrid solution
with parameters $\sin^22\theta_{e\mu}=0.5$,
 $\sin^22\theta_{e\tau}=6.0 \cdot10^{-4}$, $\Delta m^2_{31}=8.0 \cdot
10^{-6}\rm{eV}^2$ and different values of $\Delta m^2_{21}$ indicated
in the figures.}
\end{figure}

\end{document}